\documentclass[journal,12pt,onecolumn,draftclsnofoot]{IEEEtran}

\IEEEoverridecommandlockouts

\usepackage{pgfplots}
\usepackage[bookmarks,colorlinks]{hyperref}
\usepackage[linesnumbered,ruled,lined]{algorithm2e}
\usepackage{enumitem}
\usepackage{algpseudocode}
\usetikzlibrary{shapes.multipart,intersections}
\usepackage{cite}
\usepackage{amsmath,amssymb,amsfonts,amsthm,steinmetz}
\usepackage{graphicx}
\usepackage{mathrsfs}  
\usepackage{textcomp}
\usepackage{acronym}
\usepackage{xcolor}
\usepackage{upgreek,xspace}

\usepackage{tikz}
\usetikzlibrary{calc}
\makeatletter
\newcommand{\gettikzxy}[3]{%
  \tikz@scan@one@point\pgfutil@firstofone#1\relax
  \edef#2{\the\pgf@x}%
  \edef#3{\the\pgf@y}%
}
\makeatother

\usepackage[draft]{todonotes}

\usepackage{esvect}

\usepackage{times}
\usepackage{bm}
\usepackage{amsmath}
\usepackage{amssymb}
\usepackage{stmaryrd}
\usepackage{babel}
\usepackage{graphics, graphicx}
\usepackage{xcolor}
\usepackage{gensymb}
\usepackage{cite}
\usepackage{enumitem}
\usepackage{url}
\newtheorem{rem}{Remark}
\newtheorem{lem}{Lemma}

\begin{document}
\title{On the Tacit Linearity Assumption \\in Common Cascaded Models \\of RIS-Parametrized Wireless Channels}

\author{Antonin Rabault, Luc Le Magoarou, J\'{e}r\^{o}me Sol, George C. Alexandropoulos,~\IEEEmembership{Senior~Member,~IEEE}, Nir Shlezinger,~\IEEEmembership{Member,~IEEE}, \\H. Vincent Poor,~\IEEEmembership{Life Fellow,~IEEE}, and Philipp del Hougne,~\IEEEmembership{Member,~IEEE}

\thanks{Parts of this work were presented at the 17th European Conference on Antennas and Propagation (EuCAP) 2023~\cite{AntoninEuCAP}.}
\thanks{P.d.H. acknowledges funding from the CNRS pr\'{e}maturation program (project ``MetaFilt''), the Rennes M\'{e}tropole acquisition d'\'{e}quipements scientifiques program (project ``SRI''), the European Union's European Regional Development Fund, and the French region of Brittany and Rennes Métropole through the contrats de plan État-Région program (project ``SOPHIE/STIC \& Ondes'').
}
\thanks{
A. Rabault and P. del Hougne are with Univ Rennes, CNRS, IETR - UMR 6164, F-35000, Rennes, France (e-mail: \{antonin.rabault; philipp.del-hougne\}@univ-rennes1.fr).
}
\thanks{
L. Le Magoarou and J. Sol are with INSA Rennes, CNRS, IETR - UMR 6164, F-35000, Rennes, France (e-mail: \{luc.le-magoarou; jerome.sol\}@insa-rennes.fr).
}
\thanks{G. C. Alexandropoulos is with the Department of Informatics and Telecommunications,
National and Kapodistrian University of Athens, 15784 Athens, Greece (e-mail: alexandg@di.uoa.gr).
 }

 \thanks{
N. Shlezinger is with the School of ECE, Ben-Gurion University of the Negev, Beer-Sheva, Israel (e-mail: nirshl@bgu.ac.il).}

 \thanks{H. Vincent Poor is with
the Department of Electrical and Computer Engineering, Princeton University, Princeton, NJ 08544 USA (e-mail: poor@princeton.edu).}

\thanks{\textit{(Corresponding Author: Philipp del Hougne.)}
}

\vspace{-0.95cm}
}

\maketitle
	\pagestyle{plain}
	\thispagestyle{plain}
\begin{abstract}


We analytically derive from first physical principles the functional dependence of wireless channels on the RIS configuration for generic (i.e., potentially complex-scattering) RIS-parametrized radio environments. 
The wireless channel is a linear input-output relation that depends non-linearly on the RIS configuration because of two independent mechanisms: \textit{i)} proximity-induced mutual coupling between close-by RIS elements; \textit{ii)} reverberation-induced long-range coupling between all RIS elements. 
Mathematically, this ``structural'' non-linearity originates from the inversion of an ``interaction'' matrix that can be cast as the sum of an infinite Born series [for \textit{i)}] or Born-like series [for \textit{ii)}] whose $K$th term physically represents paths involving $K$ bounces between the RIS elements  [for \textit{i)}] or wireless entities [for \textit{ii)}]. 
We identify the key physical parameters that determine whether these series can be truncated after the first and second term, respectively, as tacitly done in common cascaded models of RIS-parametrized wireless channels.
Numerical results obtained with the physics-compliant PhysFad model and experimental results obtained with a RIS prototype in an anechoic (echo-free) chamber and rich-scattering reverberation chambers corroborate our analysis. 
Our findings raise doubts about the reliability of existing performance analysis and channel-estimation protocols for cases in which cascaded models poorly describe the physical reality.

\end{abstract}


\bigskip

\section{Introduction}

The performance of wireless communications systems assisted by reconfigurable intelligent surfaces (RISs)~\cite{subrt2012intelligent,Liaskos_Visionary_2018,di2019smart} is to date predominantly studied based on cascaded channel models that assume by construction a \textit{linear} parametrization of the wireless channel through the RIS. In other words, it is assumed that the wireless channel depends linearly on the RIS configuration. This practice, as we will show in this paper, is equivalent to a truncation of two infinite matrix power series (a Born series~\cite{born1926quantenmechanik,born1999principles} and a Born-like series) after the first and second term, respectively, physically meaning that any ray whose trajectory encounters more than one RIS element is ignored. A fully physics-compliant channel model would include the higher-order non-linear terms of the series, i.e., the trajectories that involve encounters with multiple RIS elements. Therefore, a physics-compliant end-to-end channel model generally depends in a \textit{non-linear} manner on the RIS configuration. The tacit linearity assumption of the commonly used linear cascaded models has to date not been justified, and the conditions under which it may (approximately) hold are still unknown. In this paper, we address these foundational questions for the modelling of RIS-parametrized wireless channels based on rigorous analytical calculations as well as numerical and experimental evidence. 

The output signal of any linear scattering system (here: the radio environment) depends linearly on the input signals via the system's transfer function (here: the wireless channel). Yet, the transfer function itself depends, in general, in a \textit{non-linear} fashion on the scattering system's structural parameters (here: the RIS configuration) because the system's response at a given location to an electromagnetic excitation is, in general, ``non-local'', i.e., it depends not only on the system's local properties but also on those at other locations. 
The concept of RIS-enabled smart radio environments constitutes a beyond-Shannon paradigm shift because, in addition to the previously available control over the input signals, RISs now yield structural control over the scattering system and hence its transfer function. 
Previously, wireless system engineers were not confronted with the non-linearity of structural parametrization since they only had control over the input signals. Now, the expansion of the available control from the input signals to both the input signals and some of the system's structural parameters inevitably entails a transition to a generally non-linear dependence of the output signals on some of the available control knobs (the structural parameters). 
The problem of the non-linearity of structural parametrization is hence intimately linked to the reason why the use of RISs in ``smart radio environments'' constitutes a paradigm shift. 

Recently, some of the current authors (and coworkers) introduced a fully physics-compliant end-to-end model for generic RIS-parametrized wireless channels: PhysFad~\cite{PhysFad}. 
The radio environment constitutes a linear time-invariant electrodynamical system (time-invariant relative to the scale of the wave's period); hence, there must be a linear operator describing the link between the incident electromagnetic fields and the  polarization fields that they induce in the system. For simplicity of notation and implementation, PhysFad assumes a sufficiently high-resolution discretization of the scattering system (i.e., the radio environment) and a dipolar scattering response of each discretized polarizable object. 
PhysFad is hence derived from first principles and describes all wireless entities (transmitters, receivers, RIS elements, scattering environment) as dipoles or collection of dipoles. The functional dependence of the wireless channel on the RIS configuration is generally non-linear in PhysFad because it involves the inversion of an ``interaction'' matrix into which the RIS configuration is encoded. It turns out, as we show in this paper, that this matrix inversion compactly captures all multi-bounce trajectories of the Born series and Born-like series, including those involving encounters with multiple RIS elements. 

Through a combination of block matrix inversion and power series expansions, applied multiple times in a hierarchical manner, we analytically derive in the present paper the Born-like series for RIS-parametrized channels from the original PhysFad formulation. Physically, the different terms of the series correspond to the different orders of multiple-scattering events. This correspondence can also be understood using graph theory, by interpreting the interaction matrix as a graph whose vertices and edges represent the dipoles and their interactions, respectively. We identify two mechanisms that must be expected to give rise to a non-linear RIS-parametrization of wireless channels in practical scenarios: \textit{i)} proximity-induced mutual coupling between the RIS elements, and \textit{ii)} reverberation-induced long-range coupling between the RIS elements. We further identify the factors that determine the importance of these mechanisms. For the former, the spatial arrangement, number and scattering strength of RIS elements matters. For the latter, the number of times that a typical ray bounces off the RIS on its trajectories from the transmitter to the receiver matters, which depends on the wave's reverberation time and the dominance of the RIS in the radio environment (i.e., the percentage of surface area covered by the RIS and the RIS elements' scattering strength). We introduce an easy-to-evaluate linearity metric to quantify to what extent the RIS parametrization of a wireless channel is linear, and we use this metric to validate our findings both numerically based on PhysFad and experimentally based on a RIS prototype.

This paper is organized as follows. We begin by discussing some generalities in Sec.~\ref{sec:Generalities} and briefly reviewing the PhysFad formalism in Sec.~\ref{sec:PhysFad}. Then, we detail our hierarchical analysis of PhysFad: We briefly analyze an antenna array in free space (Sec.~\ref{sec0}) and multiple-input multiple-output (MIMO) communications between two arrays in free space (Sec.~\ref{sec:mimo_freespace}) and go on to study RIS-assisted MIMO communication, first in free space (Sec.~\ref{sec:freespaceRIS}) and then in generic, arbitrarily complex radio environments (Sec.~\ref{sec:genericRIS}). The latter two sections include numerical and experimental evidence. Finally, we provide concluding remarks in Sec.~\ref{sec:Conclusion}, discussing in particular the consequences of the non-linearity of a generic wireless channel's dependence on the RIS configuration for channel estimation and the operation of self-adaptive RISs.

\section{Generalities}\label{sec:Generalities}

Throughout this work, we deal with linear scattering systems, meaning that, irrespective of their complexity, their input-output relation (transfer function) is linear: 
\begin{equation}
  \mathbf{y}=\mathbf{H}\mathbf{x},  
  \label{eq:linear_system}
\end{equation}
where $\mathbf{x} \in \mathbb{C}^{N_{\mathrm{T}}}$ are the input signals radiated by the $N_{\mathrm{T}}$ transmitting antennas (TX), $\mathbf{y}  \in \mathbb{C}^{N_{\mathrm{R}}}$ are the output signals captured by the $N_{\mathrm{R}}$ receiving antennas (RX), and $\mathbf{H}\in \mathbb{C}^{N_{\mathrm{R}}\times N_{\mathrm{T}}}$ is the system's transfer function.\footnote{For simplicity of exposition, we do not include a noise term in Eq.~(\ref{eq:linear_system}).} Our system of interest is a radio propagation environment equipped with one or multiple RISs. The configuration $\mathbf{c} \in \mathbb{C}^{N_{\text{S}}}$ of the $N_{\mathrm{S}}$-element RIS parametrizes the system's transfer function: $\mathbf{H}=f(\mathbf{c})$.  

Currently common channel models of RIS-parametrized radio environments postulate by construction that the parametrization function $f$ is linear with respect to $\mathbf{c}$.\footnote{To be precise, cascaded models assume that $f$ is an affine rather than linear function because $f$ may include a constant term. For simplicity, throughout this paper, we use the terminology ``linear'' irrespective of whether the constant term is zero or not.} Specifically, cascaded models work with an approximation $\hat{\mathbf{H}}$ of the physics-compliant end-to-end channel $\mathbf{H}$ of the following form:
\begin{equation}
\hat{\mathbf{H}} = \hat{f}(\mathbf{c}) = \mathbf{H}_0 + \mathbf{H}_1 \mathrm{diag}(\mathbf{c}) \mathbf{H}_2, 
\label{eq:cascade}
\end{equation}
where $\hat{\mathbf{H}}, \mathbf{H_0} \in \mathbb{C}^{N_{\mathrm{R}}\times N_{\mathrm{T}}}$, $\mathbf{H}_1 \in \mathbb{C}^{N_{\mathrm{R}}\times N_{\text{S}}}$,  and $\mathbf{H}_2 \in \mathbb{C}^{N_{\text{S}}\times N_{\mathrm{T}}}$.
The cascaded channel model from Eq.~(\ref{eq:cascade}) postulates that the end-to-end channel can be decomposed in a cascaded fashion where $\mathbf{\Phi} = \mathrm{diag}(\mathbf{c})$ captures the wavefront manipulation by the RIS and $\mathbf{H_0}$, $\mathbf{H_1}$ and $\mathbf{H_2}$ describe the wave propagation from the TX to the RX, from the RIS to the RX, and from the TX to the RIS, respectively.

To quantify the extent to which a cascaded model $\hat{\mathbf{H}}$ can describe the physical reality $\mathbf{H}$, we introduce a linearity metric $\zeta$. For simplicity, we consider a SISO case ($N_{\mathrm{T}} = N_{\mathrm{R}} = 1$), such that Eq.~(\ref{eq:cascade}) reduces to
\begin{equation}
\hat{h} = \hat{f}(\mathbf{c}) = h_0 + \mathbf{h_1}^T \mathrm{diag}(\mathbf{c}) \mathbf{h_2} = h_0 + (\mathbf{h_1}\odot\mathbf{h_2})^T \mathbf{c} = h_0 + \mathbf{t}^T \mathbf{c}, 
\label{eq:cascadeSISO}
\end{equation}
where $\mathbf{t} \triangleq (\mathbf{h_1}\odot\mathbf{h_2})$. Given the best possible choice of $h_0$ and $\mathbf{t}$ for a given setting, we determine our linearity metric as follows:
\begin{equation}
    \zeta = \frac{\mathrm{SD}_i(h_i)}{\mathrm{SD}_i(h_i-\hat{h_i})},
    \label{eq_zeta_definition}
\end{equation}
where $\mathrm{SD}_i$ denotes the standard deviation across the entries from a test data set. This definition of our linearity metric resembles the common signal-to-noise ratio (SNR) and has the advantage of being independent of the constant term (unlike the normalized mean square error). 

In order to evaluate our linearity metric in a given numerical or experimental setting, we measure two data sets (for calibration and testing), each consisting of $n$ pairs $\{ \mathbf{c}_i, h_i \}$, i.e., a random RIS configuration $\mathbf{c}_i$ and the corresponding measured channel $h_i$. We use $n=5N_{\mathrm{S}}$ for calibration and $n=100$ for testing. We perform multiple linear regression on the calibration data set in order to retrieve the parameters $h_0$ and $\mathbf{t}$ of the cascaded model that best describes the current setting. Since we consider a 1-bit programmable RIS, we fix (without loss of generality) the two possible values that the entries of the $N_{\mathrm{S}}$-element vector $\mathbf{c}$ can take to $\pm1$. Given $h_0$ and $\mathbf{t}$, we predict the channels $\hat{h_i}$ expected for the test RIS configurations and compute $\zeta$ according to Eq.~(\ref{eq_zeta_definition}).

\section{PhysFad Formulation}\label{sec:PhysFad}

\begin{figure*}
    \centering
    \includegraphics[width=\columnwidth]{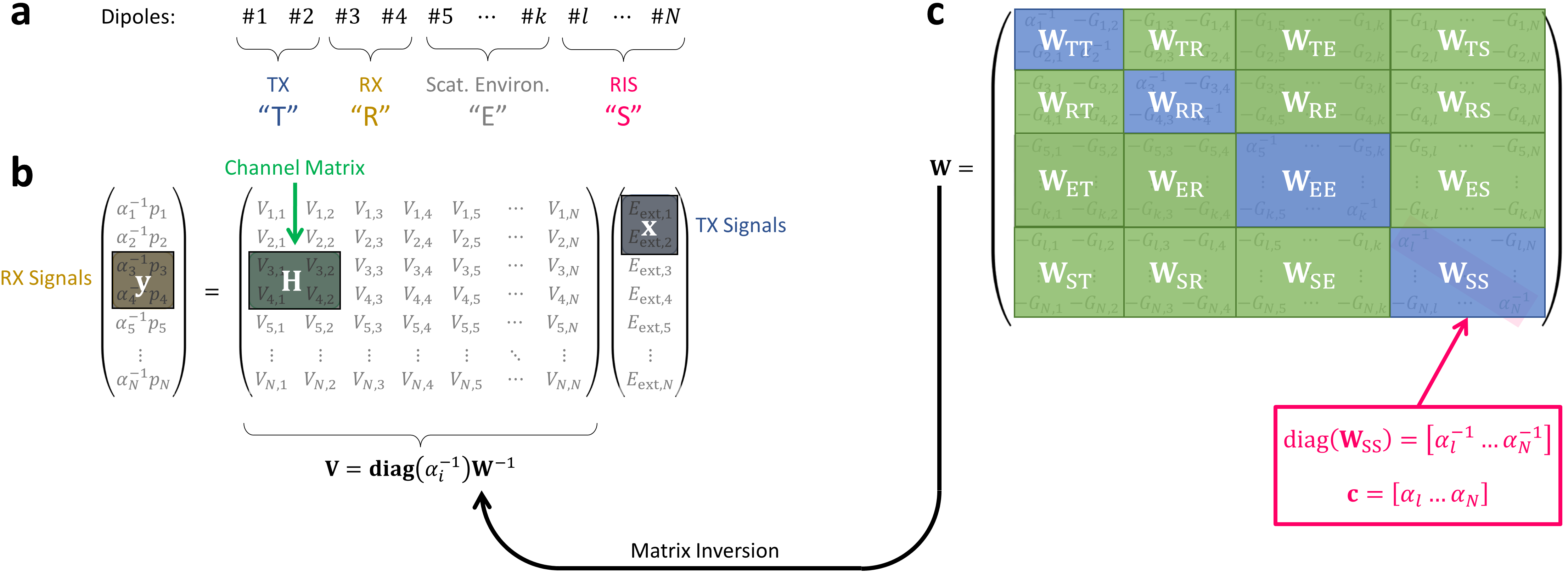}
    \caption{PhysFad formulation illustrated for a $2 \times 2$ MIMO system (adapted from Fig.~4 of Ref.~\cite{PhysFad}). a) Order of dipole indexing: T, R, E, S, where $k=N_\mathrm{T}+N_\mathrm{R}+N_\mathrm{E}$ and $l=k+1$. b) Extraction of end-to-end channel matrix $\mathbf{H}$ from $\mathbf{V}$: $\mathbf{H} = [\mathbf{V}]_{\mathrm{RT}}$. c) Block representation of $\mathbf{W}$. The relation between the diagonal of $\mathbf{W_{\mathrm{SS}}}$ and the RIS configuration $\mathbf{c}$ is also indicated. }
    \label{fig:PhysFad_basics}
\end{figure*}

For the reader's convenience and reference, we briefly summarize the essential aspects of the PhysFad formulation here, in order to prepare the ground for our subsequent analysis. For detailed explanations and derivations, the reader is referred to Ref.~\cite{PhysFad}. The wireless systems we are concerned with are composed of four wireless entities: transmitting antennas (T), receiving antennas (R), a scattering environment (E), and the RIS (S). As stated above, PhysFad describes each wireless entity as a dipole or a collection of dipoles. In total, $N = N_\mathrm{T} + N_\mathrm{R} +N_\mathrm{E} + N_\mathrm{S}$ dipoles are involved. The $i$th dipole is characterized by its polarizability $\alpha_i$ (which quantifies the dipole's tendency to acquire a dipole moment in the presence of an applied electromagnetic field) and interacts with the $j$th dipole via the free-space Green's function $G_{ij}$. The dipole's polarizability depends, among other parameters, on its resonance frequency. The simplest model of a 1-bit programmable RIS element is a single dipole that is resonant or not at the operating frequency. Each transmitter radiates an electromagnetic field, which induces dipole moments in all other dipoles, which then in turn radiate fields, etc. These interactions are captured in the interaction matrix $\mathbf{W} \in \mathbb{C}^{N\times N}$ which can be understood as consisting of $4\times 4$ blocks: T, R, E and S -- see Fig.~\ref{fig:PhysFad_basics}(c). The diagonal blocks contain the inverse polarizabilities of the corresponding dipoles along their diagonal. All other entries are the corresponding free-space Green's functions. Note in particular that the diagonal of the block indexed S (fourth row and fourth column) contains the inverse values of the polarizabilities of the RIS dipoles, i.e., the inverse values of the entries of $\mathbf{c}$ (the RIS configuration vector). The inverse of $\mathbf{W}$ is proportional to $\mathbf{V} \in \mathbb{C}^{N\times N}$, again a $4\times 4$ block matrix, whose block $[\mathbf{V}]_{\mathrm{RT}}$ (second row and first column of $\mathbf{V}$, linking the transmitting dipoles to the receiving dipoles) is the physics-compliant end-to-end channel matrix $\mathbf{H}$ -- see Fig.~\ref{fig:PhysFad_basics}(b). The dependence of $\mathbf{H}$ on $\mathbf{c}$ is hence generally non-linear.

A dipole is an isotropic scattering object and/or radiator. To capture anisotropic scattering properties, a single dipole can be replaced by a collection of dipoles whose various dipole parameters have been optimized such that they collectively have the desired anisotropic characteristics~\cite{bertrand2020global}. In the PhysFad formulation, a single dipole entry of $\mathbf{W}$ is then accordingly replaced by a block of dipole entries corresponding to this collection of dipoles. 

Finally, note that for simplicity of notation and exposition, PhysFad was introduced for 2D geometries in Ref.~\cite{PhysFad}. We follow this version in the present paper. An extension to a dyadic 3D formulation is conceptually straightforward.
Moreover, to simplify the notation, PhysFad is formulated in a dimensionless unit system.

Prior to PhysFad, it had already been noticed that common signal-processing models of communications systems are in general \textit{not} guaranteed to be consistent with the physical laws governing their corresponding experimental realizations, leading to the proposal of a multiport circuit theory of communications systems~\cite{ivrlavc2010toward,ivrlavc2014multiport}. These works took mutual coupling between the elements of antenna arrays into account, but they did not consider the possibility of RIS-parametrized wave propagation environments (including potentially highly complex scattering structures). More recently, mutual coupling between the RIS elements was modelled in Refs.~\cite{williams2020communication,gradoni_EndtoEnd_2020,badheka2023accurate}, however under the strongly limiting assumptions of operation in free space with minimally scattering antenna arrays. Moreover, no exact understanding of the non-linear dependence of the end-to-end channel matrix on the RIS configuration had been worked out; in the present paper, the Born series for mutual coupling between the RIS elements in Eq.~(\ref{eq:wss_series}) elucidates the nature of this non-linearity. A general fully physics-compliant framework for the analysis of RIS-parametrized wireless channels in generic (potentially highly complex) scattering environments has been introduced only recently with PhysFad~\cite{PhysFad}. In particular, the reverberation-induced long-range coupling between the RIS elements that we highlight in Sec.~\ref{sec:genericRIS} has been completely overlooked in the recent signal-processing literature (apart from some recent works by one of the current authors and coworkers~\cite{AntoninEuCAP,PhysFad,ChloeMag}).

\begin{rem}
    PhysFad does not require any \textit{ad hoc} corrections because it is derived from first principles and complies with all relevant physical laws. For instance, pathloss, frequency selectivity and the intertwinement of amplitude and phase response are all automatically accounted for.
\end{rem}

In the following sections, we apply multiple times in a hierarchical manner a block matrix inversion and power series expansion to the interaction matrix $\mathbf{W}$ in order to study the functional dependence of $\mathbf{H} \propto [\mathbf{W}^{-1}]_\mathrm{RT}$ on $\mathbf{c}$. We will see that the resulting infinite series of terms represent multi-bounce trajectories between scattering entities that are compactly captured through the matrix inversion in PhysFad and responsible for the ``structural non-linearity'', i.e., the non-linear dependence of the linear wireless channel on the RIS configuration.

\section{One Antenna Array in Free Space}
\label{sec0}

To start, let us consider an antenna array in free space. This very simple scenario is the first step of our hierarchical analysis. Of course, with a single antenna array in free space, there is not yet any notion of communication and no channel matrix between TX and RX can be defined. $\mathbf{W}$ is simply equal to $\mathbf{W}_{\mathrm{TT}}$ in this case and $\mathbf{W}_{\mathrm{TT}}^{-1}$ will be proportional to the fields that the transmitting antenna array induces on itself. 
We can decompose $\mathbf{W}_{\mathrm{TT}} = \mathbf{\Omega}_{\mathrm{TT}}^{-1} + \mathbf{\mathcal{M}}_{\mathrm{TT}}$, where $\mathbf{\Omega}_{\mathrm{TT}}^{-1} = \mathrm{diag}[\alpha_1^{-1} \ \dots \ \alpha_{N_\text{T}}^{-1} ]$ contains the inverse polarizabilities of the antenna dipoles, and the off-diagonal elements of the matrix $\mathbf{\mathcal{M}}_{\mathrm{TT}}$ contain the free-space Green's functions between the antenna elements (the diagonal elements of $\mathbf{\mathcal{M}}_{\mathrm{TT}}$ are zero).
Now, neglecting mutual coupling between the antennas amounts to setting $\mathbf{\mathcal{M}}_{\mathrm{TT}}=\mathbf{0}$ and 
one obtains $\mathbf{W}_{\mathrm{TT}}^{-1} = \mathbf{\Omega}_{\mathrm{TT}} = \mathrm{diag}[\alpha_1 \ \dots \ \alpha_{N_\text{T}} ]$. The physics-compliant result without neglecting mutual coupling is 
\begin{equation}
\mathbf{W}_{\mathrm{TT}}^{-1} = \left( \mathbf{\Omega}_{\mathrm{TT}}^{-1} + \mathbf{\mathcal{M}}_{\mathrm{TT}} \right)^{-1} = \left( \mathbf{I} + \mathbf{\Omega}_{\mathrm{TT}} \mathbf{\mathcal{M}}_{\mathrm{TT}} \right)^{-1} \mathbf{\Omega}_{\mathrm{TT}}.
\label{eq8}
\end{equation}
We can express the inverse of $\mathbf{W}_{\mathrm{TT}}$ now as an infinite power series: 
\begin{equation}
\mathbf{W}_{\mathrm{TT}}^{-1} = \left(\sum_{k=0}^\infty 
\left( -\mathbf{\Omega} _{\mathrm{TT}}\mathbf{\mathcal{M}}_{\mathrm{TT}} \right)^{k}\right) \mathbf{\Omega}_{\mathrm{TT}},
\label{eq_wttseries}
\end{equation}

\noindent and writing down the first few terms of the infinite series from Eq.~(\ref{eq_wttseries}), we obtain a Born series:
\begin{equation}
    \mathbf{W_{\mathrm{TT}}}^{-1} = \mathbf{\Omega_{\mathrm{TT}}}  - \mathbf{\Omega_{\mathrm{TT}}} \mathbf{\mathcal{M}_{\mathrm{TT}}}\mathbf{\Omega_{\mathrm{TT}}}  + \left(\mathbf{\Omega_{\mathrm{TT}}}\mathbf{\mathcal{M}_{\mathrm{TT}}}\right)^2\mathbf{\Omega_{\mathrm{TT}}} - \dots
    \label{eq:wtt_series}
\end{equation}

\noindent A Born series is the expansion of a scattering quantity in terms of the interaction potential, named after Max Born who studied particles in scattering potentials in quantum mechanics~\cite{born1926quantenmechanik,jost1951scattering}. The same formalism applies to classical electromagnetic waves~\cite{van1999multiple,born1999principles} and has long been used to study, for example, light scattering from small penetrable objects.  However, the Born series can diverge for large scattering systems and strong scattering potentials~\cite{corbett1968convergence}. Upon suitable modification of the Born series with a preconditioner, convergence of the series can be guaranteed in all cases~\cite{osnabrugge2016convergent}.

Our (unmodified) Born series converges if the spectral radius of $\mathbf{\Omega}_{\mathrm{TT}}\mathbf{\mathcal{M}}_{\mathrm{TT}}$ is below unity: $\rho (\mathbf{\Omega}_{\mathrm{TT}} \mathbf{\mathcal{M}}_{\mathrm{TT}}) <1$~\cite[p.~195]{lax2002functional}.
The matrix $\mathbf{\Omega}_{\mathrm{TT}}\mathbf{\mathcal{M}}_{\mathrm{TT}}$ is the common ratio of our Born series, and its norm quantifies the attenuation each term undergoes compared to the previous one. Therefore, the smaller it is, the earlier we can truncate the series for a given desired error level. 
Assuming that all antennas have the same polarizability $\alpha_\text{T}$, we show in Appendix~\ref{AppendixA} that
\begin{equation}
    \left\Vert\mathbf{\Omega}_{\mathrm{TT}} \mathbf{\mathcal{M}}_{\mathrm{TT}} \right\Vert_2 \leq \mathcal{C}_\text{T} \triangleq |\alpha_\text{T}|\ \underset{i \in [1,N_\text{T}]}{\text{max}}\,  \sum_{\substack{j\in [1,N_\text{T}] \\ j \neq i}} |G_{ij}|.
    \label{definition_C}
\end{equation}
Hence, the Born series converges if $\mathcal{C}_\text{T}<1$. 
The term $\mathcal{C}_\text{T}$ is a dimensionless characteristic of the antenna array that is proportional to the magnitude of the antenna polarizability and increases as the spacing between antennas is reduced. It follows (see Appendix~\ref{AppendixA}) that the normalized mean square error due to truncating the series after the $K$th term is bounded by
\begin{equation}
     \mathcal{C}_\text{T}^K\frac{1+\mathcal{C}_\text{T}}{1-\mathcal{C}_\text{T}}.
     \label{bound_antenna_array}
\end{equation}

\section{MIMO Communications in Free Space}
\label{sec:mimo_freespace}

Next, let us consider wireless MIMO communications between a TX array and an RX array in free space without a RIS. This simple scenario is the second step of our hierarchical analysis. Because there is no scattering environment or RIS, $\mathbf{W}$ here simply takes the form of

\begin{equation}
\mathbf{W}_1 = 
    \begin{bmatrix} 
	\mathbf{W}_{\text{TT}} & \mathbf{W}_{\text{TR}} \\
	\mathbf{W}_{\text{RT}} & \mathbf{W}_{\text{RR}}\\
	\end{bmatrix}.
 \label{eq:W1}
\end{equation}

The $N_\mathrm{R}\times N_\mathrm{T}$ channel matrix $\mathbf{H}$ between the TX array and the RX array is proportional to the bottom left block of $\mathbf{W}_1^{-1}$, i.e., $\mathbf{H} \varpropto [\mathbf{W}_1^{-1}]_{\mathrm{RT}}$. Standard formulas for the inversion of a block matrix like $\mathbf{W}_1$ yield

\begin{equation}
\begin{split}
    [\mathbf{W}_1^{-1}]_{\mathrm{RT}} = - \mathbf{W}_{\text{RR}}^{-1} \mathbf{W}_{\text{RT}} \left( \mathbf{W}_{\text{TT}} - \mathbf{W}_{\text{TR}} \mathbf{W}_{\text{RR}}^{-1} \mathbf{W}_{\text{RT}} \right)^{-1} 
    \\= - \mathbf{W}_{\text{RR}}^{-1} \mathbf{W}_{\text{RT}} \mathbf{W}_{\text{TT}}^{-1} \left( \mathbf{I} - \mathbf{W}_{\text{TR}} \mathbf{W}_{\text{RR}}^{-1} \mathbf{W}_{\text{RT}} \mathbf{W}_{\text{TT}}^{-1} \right)^{-1} 
    \\= - \mathbf{W}_{\text{RR}}^{-1} \mathbf{W}_{\text{RT}} \mathbf{W}_{\text{TT}}^{-1} \sum_{k=0}^\infty \left( \mathbf{W}_{\text{TR}} \mathbf{W}_{\text{RR}}^{-1} \mathbf{W}_{\text{RT}} \mathbf{W}_{\text{TT}}^{-1} \right)^k
    \\ = - \mathbf{W}_{\text{RR}}^{-1} \mathbf{W}_{\text{RT}} \mathbf{W}_{\text{TT}}^{-1} - \mathbf{W}_{\text{RR}}^{-1} \mathbf{W}_{\text{RT}} \mathbf{W}_{\text{TT}}^{-1} \mathbf{W}_{\text{TR}} \mathbf{W}_{\text{RR}}^{-1} \mathbf{W}_{\text{RT}} \mathbf{W}_{\text{TT}}^{-1}  - \dots.
\end{split}
\label{eq_w1inv}
\end{equation}

\begin{rem}
Similar to Eq.~(\ref{eq_wttseries}) from the previous section, we have a power series in Eq.~(\ref{eq_w1inv}). However, the structure of the matrix whose powers we add is different here, and Eq.~(\ref{eq_wttseries}) does not constitute a Born series. Hence, we refer to Eq.~(\ref{eq_w1inv}) as a Born-like series.
\end{rem}

The series in Eq.~(\ref{eq_w1inv}) converges if $\rho\left( \mathbf{W}_{\text{TR}} \mathbf{W}_{\text{RR}}^{-1} \mathbf{W}_{\mathrm{RT}} \mathbf{W}_{\mathrm{TT}}^{-1} \right)<1$, and a fortiori if 

\noindent $\left\Vert \mathbf{W}_{\text{TR}} \mathbf{W}_{\text{RR}}^{-1} \mathbf{W}_{\mathrm{RT}} \mathbf{W}_{\mathrm{TT}}^{-1} \right\Vert_2<1$. We show in Appendix~\ref{AppendixB} that this norm can be bounded as 
\begin{equation}
\left\Vert \mathbf{W}_{\text{TR}} \mathbf{W}_{\text{RR}}^{-1} \mathbf{W}_{\mathrm{RT}} \mathbf{W}_{\mathrm{TT}}^{-1} \right\Vert_2 \leq \frac{|\alpha_\text{T}|}{1-\mathcal{C}_\text{T}} \frac{|\alpha_\text{R}|}{1-\mathcal{C}_\text{R}}N_{\text{T}}N_{\text{R}} \frac{k^2}{4\epsilon\delta} \left| \mathrm{H}_0^{(2)}(kD_{\text{RT}}) \right|^2,
\label{eq:WTRRRRTTT_bound}
\end{equation}
where $D_{\mathrm{RT}}$ denotes the smallest distance between any transmitter-receiver antenna pair, and $\alpha_\text{R}$ and $\mathcal{C}_\text{R}$ are characteristics of the RX array defined analogous to $\alpha_\text{T}$ and $\mathcal{C}_\text{T}$ for the TX array (see Sec.~\ref{sec0}). Since the separation between the TX array and the RX array is generally many wavelengths large, the common ratio of the infinite series in Eq.~(\ref{eq_w1inv}) is generally very small and subsequent terms of the series are attenuated at least by a factor $O(D_{\text{RT}}^k)$ compared to the first one. However, it is difficult to obtain a tight bound on the error due to truncating the series because $\mathbf{W}_{\text{RT}}=\mathbf{W}_{\text{TR}}^T$ is generally very ill-conditioned (or even of rank unity). 

The first term of the Born-like series in Eq.~(\ref{eq_w1inv}) contains the direct TX-RX path. The $K$th term represents $K$ round trips (TX-RX-TX) followed by a direct TX-RX path. Thus, the common ratio of the infinite series represents a round trip between the TX array and the RX array.

\begin{rem}
    The first term of the Born-like series in Eq.~(\ref{eq_w1inv})   is not trivially the free-space transmission matrix $\mathbf{W}_{\mathrm{RT}}$ due to the mutual coupling present in each antenna array that we discussed in the previous Sec.~\ref{sec0}.
\end{rem}

The first term winds up being directly proportional to $\mathbf{W}_{\text{RT}}$ only under Assumption~1:

\textit{Assumption 1:} All antenna elements are identical (i.e., have the same polarizability) and there is no mutual coupling between the elements of a given antenna array. Then, $\mathbf{W}_{\mathrm{RR}}^{-1}$ and $\mathbf{W}_{\mathrm{TT}}^{-1}$ are scaled identity matrices.

The higher-order terms in Eq.~(\ref{eq_w1inv}) involve round trips before being captured by the RX array and can be ignored under Assumption~2:

\textit{Assumption 2:} The antenna elements have small scattering cross-sections (i.e., $\lVert \mathbf{W}_{\mathrm{RR}} \rVert_2$ and $\lVert \mathbf{W}_{\mathrm{TT}} \rVert_2$ are small) and/or the TX array and the RX array are far apart (i.e., $\lVert \mathbf{W} _{\mathrm{TR}}\rVert_2=\lVert \mathbf{W}_{\mathrm{RT}} \rVert_2$ is small). The assumption of vanishing scattering cross-section is related to the concept of ``canonical minimum scattering antenna``~\cite{kahn1965minimum,ivrlavc2014multiport}. Examples of antennas with weak or strong scattering cross-section are a short dipole and a horn antenna, respectively.

For later reference, at this stage we can also work out the series expansions of $ [\mathbf{W}_1^{-1}]_{\mathrm{TT}}$ and $ [\mathbf{W}_1^{-1}]_{\mathrm{RR}}$, which, under Assumption~2, turn out to be equal to $\mathbf{W}_{\mathrm{TT}}^{-1}$ and $\mathbf{W}_{\mathrm{RR}}^{-1}$, respectively.

\section{RIS-Assisted MIMO Communications in Free Space}
\label{sec:freespaceRIS}

We are now all set to consider RIS-assisted MIMO communications between a TX array and an RX array \textit{in free space}. This scenario is the third step of our hierarchical analysis. 
In this section, we begin with an analytical analysis  (Sec.~\ref{freespaceRIS_analytical}) similar to the previous two sections, followed by a numerical analysis (Sec.~\ref{freespaceRIS_numerical}) and an experimental analysis (Sec.~\ref{freespaceRIS_experimental}).

\subsection{Analytical Analysis}
\label{freespaceRIS_analytical}

Because we assume to operate in free space, there is no scattering environment and $\mathbf{W}$ here takes the form of

\begin{equation}
\mathbf{W}_2 = 
    \begin{bmatrix} 
	\mathbf{W}_{\mathrm{TT}} & \mathbf{W}_{\mathrm{TR}} & \mathbf{W}_{\mathrm{TS}} \\
	\mathbf{W}_{\mathrm{RT}} & \mathbf{W}_{\mathrm{RR}} & \mathbf{W}_{\mathrm{RS}}\\
	\mathbf{W}_{\mathrm{ST}} & \mathbf{W}_{\mathrm{SR}} & \mathbf{W}_{\mathrm{SS}}\\
	\end{bmatrix}
 =  \begin{bmatrix} 
	\mathbf{W}_1  & \mathbf{W}_{\mathrm{1S}} \\
	\mathbf{W}_{\mathrm{S1}}  & \mathbf{W}_{\mathrm{SS}} \\
\end{bmatrix},
 \label{eq:W2}
\end{equation}

\noindent where $\mathbf{W}_{\mathrm{S1}} = [\mathbf{W}_{\mathrm{ST}} \ \mathbf{W}_{\mathrm{SR}}]$ and $\mathbf{W}_{\mathrm{1S}} = [\mathbf{W}_{\mathrm{TS}} \ \mathbf{W}_{\mathrm{RS}}]^T$.

The $N_\mathrm{R}\times N_\mathrm{T}$ channel matrix $\mathbf{H}$ between the TX array and the RX array is proportional to the middle left block of $\mathbf{W}_2^{-1}$, i.e., $\mathbf{H} \varpropto [\mathbf{W}_2^{-1}]_{\mathrm{RT}}$. Standard formulas for the inversion of a block matrix like $\mathbf{W}_2$ yield

\begin{equation}
\begin{split}
    [\mathbf{W}_2^{-1}]_{1} = \left( \mathbf{W}_1 - \mathbf{W}_{\mathrm{1S}} \mathbf{W}_{\mathrm{SS}}^{-1} \mathbf{W}_{\mathrm{S1}} \right)^{-1} 
    \\=  \mathbf{W}_{1}^{-1} \left( \mathbf{I} - \mathbf{W}_{\mathrm{1S}} \mathbf{W}_{\mathrm{SS}}^{-1} \mathbf{W}_{\mathrm{S1}} \mathbf{W}_{1}^{-1} \right)^{-1} 
    =  \mathbf{W}_{1}^{-1} \sum_{k=0}^\infty \left( \mathbf{W}_{\mathrm{1S}} \mathbf{W}_{\mathrm{SS}}^{-1} \mathbf{W}_{\mathrm{S1}} \mathbf{W}_{1}^{-1} \right)^k
    \\= \mathbf{W}_{1}^{-1} + \mathbf{W}_{1}^{-1}\mathbf{W}_{\mathrm{1S}} \mathbf{W}_{\mathrm{SS}}^{-1} \mathbf{W}_{\mathrm{S1}}\mathbf{W}_{1}^{-1} + \mathbf{W}_{1}^{-1}\left(\mathbf{W}_{\mathrm{1S}} \mathbf{W}_{\mathrm{SS}}^{-1} \mathbf{W}_{\mathrm{S1}} \mathbf{W}_{1}^{-1} \right)^2 + \dots
\end{split}
\label{eq_w2inv}
\end{equation}

\noindent This is another Born-like series. It converges if $\rho\left( \mathbf{W}_{\mathrm{1S}} \mathbf{W}_{\mathrm{SS}}^{-1} \mathbf{W}_{\mathrm{S1}} \mathbf{W}_{1}^{-1} \right)<1$ and an analysis analogous to the one from Sec.~\ref{sec:mimo_freespace} leading to Eq.~(\ref{eq:WTRRRRTTT_bound}) can be performed. Subsequent terms in the series rapidly attenuate in typical settings where the TX array, the RX array and the RIS are separated from each other by many wavelengths. Under Assumption~2, we can justify truncating the Born-like series from Eq.~(\ref{eq_w2inv}) at linear order (i.e., neglecting all terms with non-linear dependence on $\mathbf{W}_{\mathrm{SS}}^{-1}$).

This time, we are not done yet because $\mathbf{H} \varpropto [\mathbf{W}_2^{-1}]_{\mathrm{RT}}$ and so far we have only worked out $[\mathbf{W}_2^{-1}]_{1}$. Using Assumption~2 and the definition of $\mathbf{W}_{\mathrm{S1}}$ and $\mathbf{W}_{\mathrm{1S}}$ yields

\begin{equation}
    [\mathbf{W}_2^{-1}]_{\mathrm{RT}} = [\mathbf{W}_{1}^{-1}]_{\mathrm{RT}} + 
    \left[ [\mathbf{W}_{1}^{-1}]_{\mathrm{RT}}\ [\mathbf{W}_{1}^{-1}]_{\mathrm{RR}}\right] \begin{bmatrix} \mathbf{W}_{\mathrm{TS}} \\ \mathbf{W}_{\mathrm{RS}}\end{bmatrix} \mathbf{W}_{\mathrm{SS}}^{-1} \ \left[\mathbf{W}_{\mathrm{ST}} \ \mathbf{W}_{\mathrm{SR}}\right] \begin{bmatrix}[\mathbf{W}_{1}^{-1}]_{\mathrm{TT}} \\ [\mathbf{W}_{1}^{-1}]_{\mathrm{RT}}\end{bmatrix}.
\label{eq_RIS_series}
\end{equation}

\noindent Comparing the expression in Eq.~(\ref{eq_RIS_series}) to the common cascaded model from Eq.~(\ref{eq:cascade}), we can identify the following correspondences:

\begin{subequations}
\begin{align}
\mathbf{H_0} \longleftrightarrow   [\mathbf{W}_{1}^{-1}]_{\mathrm{RT}}\\
\mathbf{H_1} \longleftrightarrow   [\mathbf{W}_{1}^{-1}]_{\mathrm{RT}}\mathbf{W}_{\mathrm{TS}} + [\mathbf{W}_{1}^{-1}]_{\mathrm{RR}} \mathbf{W}_{\mathrm{RS}}\\
\mathbf{H_2} \longleftrightarrow   \mathbf{W}_{\mathrm{ST}}[\mathbf{W}_{1}^{-1}]_{\mathrm{TT}}  + \mathbf{W}_{\mathrm{SR}} [\mathbf{W}_{1}^{-1}]_{\mathrm{RT}}
\end{align}
\label{eq:interpretonlyRIS}
\end{subequations}

\noindent The terms in Eq.~(\ref{eq:interpretonlyRIS}) have obvious physical interpretations. The term corresponding to $\mathbf{H_0}$ represents the family of paths between the TX array and the RX array that do not interact with the RIS. Recall that we have previously worked out $[\mathbf{W}_1^{-1}]_{\mathrm{RT}}$ at the end of Sec.~\ref{sec:mimo_freespace}; this family of paths includes those that bounce multiple times between the TX array and the RX array. Only the direct path without any such bounces corresponds to the conventional understanding of a ``line of sight'' (LOS) path. 
The term corresponding to $\mathbf{H_1}$ contains the two possible families of paths from the RIS to the RX array without encountering the RIS along the way, namely those that start with a trajectory from the RIS to the TX array or the RX array, and eventually arrive at the RX array. Those starting with a trajectory from the RIS to the TX array can be neglected under Assumption~2. 
The term corresponding to $\mathbf{H_2}$ contains the two possible families of paths from the TX array to the RIS without encountering the RIS along the way, namely those that start from the TX array, and, without any previous encounters with the RIS, finish with a trajectory from the TX array or the RX array to the RIS. Those finishing with a trajectory from the RX array to the RIS can again be neglected under Assumption~2. 

However, we have not yet arrived at the cascaded model because the (approximate) channel model in Eq.~(\ref{eq_RIS_series}) is linear in $\mathbf{W}_{\mathrm{SS}}^{-1}$ but not in $\mathbf{c}$. Akin to our course of action in Sec.~\ref{sec0}, we can decompose $\mathbf{W}_{\mathrm{SS}} = \mathbf{\Phi}^{-1} + \mathbf{\mathcal{M}}_{\mathrm{SS}}$, where $\mathbf{\Phi} = \mathrm{diag}(\mathbf{c})$ and $\mathbf{\mathcal{M}}_{\mathrm{SS}}$ captures the mutual coupling between the RIS elements. By analogy with Sec.~\ref{sec0}, it follows immediately that 

\begin{equation}
    \mathbf{W}_{\mathrm{SS}}^{-1} = \mathbf{\Phi}  + \mathbf{\Phi} \mathbf{\mathcal{M}}_{\mathrm{SS}}\mathbf{\Phi}  + \left(\mathbf{\Phi}\mathbf{\mathcal{M}}_{\mathrm{SS}}\right)^2\mathbf{\Phi} + \dots
    \label{eq:wss_series}
\end{equation}
and the common ratio of this Born series is bounded as follows:
\begin{equation}
    \left\Vert\mathbf{\Phi} \mathbf{\mathcal{M}}_{\mathrm{SS}}\right\Vert_2 \leq \mathcal{C}_\mathrm{S} \triangleq |\alpha_\text{S}|\ \underset{i \in [\eta+1,\eta+N_\text{S}]}{\text{max}}\,  \sum_{\substack{j\in [\eta+1,\eta+N_\text{S}] \\ j \neq i}} |G_{ij}|,
    \label{definition_C_RIS}
\end{equation}
where we use $\eta=N_\text{T}+N_\text{R}$ (recall $N_\text{E}=0$ in free space) for conciseness.

The cascaded channel model assumes $\mathbf{W}_{\mathrm{SS}}^{-1} = \mathbf{\Phi}$, i.e., a truncation of the Born series in Eq.~(\ref{eq:wss_series}) after the first term.  This truncation is justified if Assumption~3 holds:

\textit{Assumption 3:} The mutual coupling between the RIS elements is negligible.

A bound on the error due to Assumption~3 can be formulated analogous to Eq.~(\ref{bound_antenna_array}) from Sec.~\ref{sec0}. Future work may be able to identify tighter bounds, for example, in the case of a 1-bit programmable RIS by accounting for the fact that on average half of the RIS elements are in their ``OFF'' state, and hence have approximately zero polarizability. The error due to neglecting proximity-induced mutual coupling between the RIS elements depends on key properties of the RIS as follows:
\begin{enumerate}
    \item \textit{Scattering cross-section of RIS elements.} The scattering cross-section of the RIS elements depends on their polarizabilities and hence directly relates to $\lVert \mathbf{\Phi} \rVert_2$. The larger the scattering cross-section is, the more strongly the RIS elements will interact with their neighbors. Correspondingly, a larger scattering cross-section of the RIS elements leads to a larger value of $\lVert \mathbf{\Phi} \rVert_2$ and hence a slower convergence of the Born series in Eq.~(\ref{eq:wss_series}). 
    \item \textit{Number of RIS elements.} The more RIS elements there are, the more detrimental proximity-induced mutual-coupling effects must be expected. Both $\lVert \mathbf{\Phi} \rVert_2$ and $\lVert \mathbf{\mathcal{M}}_{ \mathrm{SS}} \rVert_2$ will be larger if there are more RIS elements. In the limiting case of $N_{\mathrm{S}}=1$, there is obviously no mutual coupling between the RIS elements and the inversion of $\mathbf{W}_{\mathrm{SS}}^{-1}$ becomes trivial. 
     \item \textit{Spatial arrangement of RIS elements.} The Green's function between the RIS elements depends on their spatial arrangement: the spacing between the RIS elements and the RIS surface topology (e.g., planar vs. curved surface). The magnitudes of the Green's functions directly impact the value of $\lVert \mathbf{\mathcal{M}}_{ \mathrm{SS}} \rVert_2$. 
\end{enumerate}

In order to minimize the error due to Assumption~3 (i.e., to achieve a free-space setting in which the cascaded channel model can be used), various possibilities arise:
\begin{enumerate}
\item \textit{Use RIS elements with small scattering cross-section.} However, this counteracts our wish that RIS elements should strongly impact the wireless channel. In the limit of vanishing scattering cross-section, the mutual coupling becomes negligible but the RIS elements become useless.
\item \textit{Use few RIS elements.} However, this counteracts our wish that RIS elements should be multitudinous~\cite{jamali2022impact}. In the limit of a single RIS element, there is no mutual coupling but we are deprived of the ability to significantly control the wireless channel.
\item \textit{Optimize the spatial arrangement of the RIS elements to minimize mutual coupling.} This is the most difficult but most promising option. The most obvious parameter to consider is the spacing of the RIS elements, but the RIS surface topology should also be accounted for. Indeed, future conformal RIS prototypes whose surface topology is adapted to non-planar surfaces may yield stronger mutual coupling than their otherwise identical planar counterparts because the curvature modifies the Green's functions between the RIS elements. Moreover, one may on purpose induce additional coupling effects such that mutual coupling is overall reduced or mitigated. Efforts to mitigate mutual coupling by adding decoupling mechanisms to the RIS are yet to be transposed from the design of conventional patch-antenna arrays~\cite{wu2017array,vishvaksenan2017mutual,li2018isolation,lin2020weak,zhang2021novel,zhang2021simple} to the design of RIS prototypes.
\end{enumerate}

\subsection{Numerical Analysis}
\label{freespaceRIS_numerical}

\begin{figure*}
    \centering
    \includegraphics[width=0.8\columnwidth]{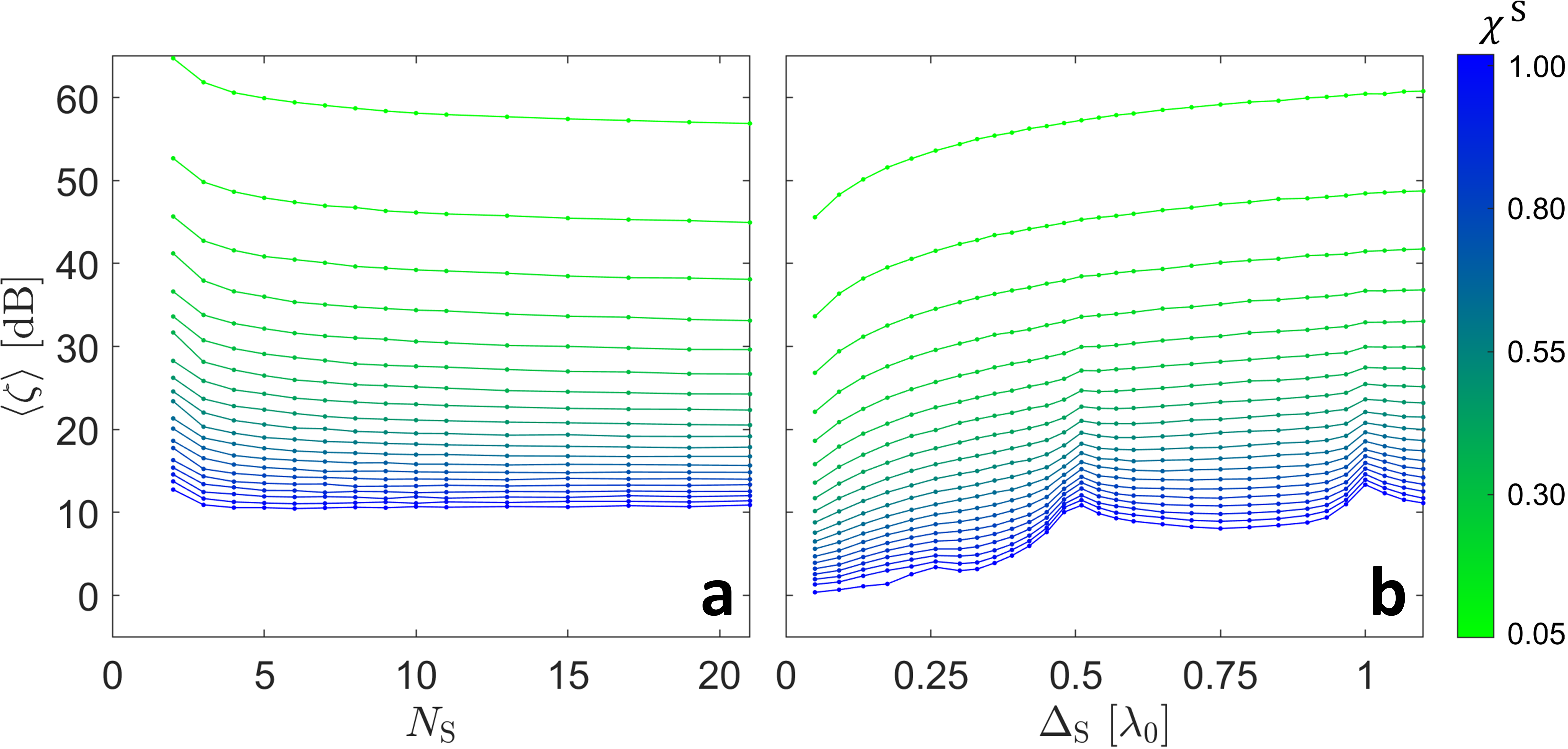}
    \caption{Numerical evaluation based on PhysFad of how the linearity metric of a SISO channel's RIS parametrization in free space depends on the RIS element's scattering cross-section (color-coded), on the number of RIS elements $N_\mathrm{S}$ (a), and on the RIS element spacing $\Delta_{\mathrm{S}}$ (b). Every shown data point is averaged over 1000 realizations with randomly chosen TX and RX positions.}
    \label{fig:RISfreespacePhysFad}
\end{figure*}

To confirm these analytical insights into the extent to which proximity-induced mutual coupling between the RIS elements limits the applicability of the linear cascaded channel model to a RIS-assisted MIMO wireless communications system \textit{in free space}, we now conduct numerical tests with PhysFad~\cite{PhysFad}. We consider a RIS-assisted SISO system in free space. The 2D PhysFad setup involves a planar 1D RIS whose elements are spaced by $\Delta_{\mathrm{S}}$; the transmitting antenna and the receiving antenna are both omnidirectional dipoles located on the same side of the RIS, at least six wavelengths away from the RIS. We evalute our linearity metric $\zeta$ (introduced in Sec.~\ref{sec:Generalities}) over 1000 random choices of TX location and RX location. 
We systematically evaluate the dependence of $\langle \zeta \rangle$ on \textit{i)} the scattering cross-section of the RIS elements (controlled via the dipole parameter $\chi_i$: $\alpha_i \propto \chi_i^2$, see Ref.~\cite{PhysFad})\footnote{The range of possible values of $\chi^{\mathrm{S}}$ has an upper bound imposed by the chosen value of $\gamma_i^R$ since energy conservation requires that $\mathrm{Im}(\alpha_i^{-1})\geq\mu$. ~\cite{PhysFad}}, \textit{ii)} the number of RIS elements  $N_{\mathrm{S}}$, and \textit{iii)} the spacing $\Delta_{\mathrm{S}}$ between the RIS elements. The parameters of the TX antenna and the RX antenna are always the same such that any dependencies of $\langle\zeta\rangle$ observed in Fig.~\ref{fig:RISfreespacePhysFad} are due to the RIS properties. The specific antenna dipole properties are $\chi^{\mathrm{T}}=\chi^{\mathrm{R}}=1$ and $f_{\mathrm{res}}^{\mathrm{T}}=f_{\mathrm{res}}^{\mathrm{R}}=1$. Moreover, throughout this paper we use $\Gamma^{L}=0$ for all dipoles. (See Ref.~\cite{PhysFad} for additional background on dipole parameters.)

Our results are shown in Fig.~\ref{fig:RISfreespacePhysFad} and confirm the expected trends discussed in the previous sub-section (Sec.~\ref{freespaceRIS_analytical}). Specifically, we make the following observations in Fig.~\ref{fig:RISfreespacePhysFad}:

\begin{enumerate}
  \item A larger scattering cross-section of the RIS elements yields a deterioration of the linearity metric, \textit{ceteris paribus}, because proximity-induced mutual coupling is more important. 
  \item A larger number of RIS elements yields a deterioration of the linearity metric, \textit{ceteris paribus}, because more mutual-coupling effects arise. However, the linearity metric only significantly depends on $N_{\mathrm{S}}$ for small values of $N_{\mathrm{S}}$. Once there are many RIS elements already, adding another one does not add significant additional proximity-induced mutual-coupling effects to most of the already existing RIS elements.
  \item The linearity metric strongly depends on $\Delta_{\mathrm{S}}$. For small values of $\chi^{\mathrm{S}}$ (color-coded in green), $\mathrm{Im(\alpha^{-1})} \gg \mu$ such that $\mathbf{W}_{\mathrm{SS}}$ has a strong diagonal and the linearity metric monotonously improves as $\Delta_{\mathrm{S}}$ is increased, because the mutual-coupling effects get weaker. For large values of $\chi^{\mathrm{S}}$ (color-coded in blue), $\mathrm{Im(\alpha^{-1})} \approx \mu$ such that the diagonal and off-diagonal entries of $\mathbf{W}_{\mathrm{SS}}$ are of a similar order of magnitude, and the dependence of the linearity metric on $\Delta_{\mathrm{S}}$ is not monotonous. Notable peaks are seen in the vicinity of  $\Delta_{\mathrm{S}} = 0.5\lambda_0$ and $\Delta_{\mathrm{S}} = \lambda_0$. We attribute these peaks to the properties of the Hankel function that is involved in the off-diagonal entries of $\mathbf{W}_{\mathrm{SS}}$. Its real-valued arguments are integer multiples of $k\Delta_{\mathrm{S}}$; while the magnitude of $\mathrm{H}_0^{(2)}(k\Delta_{\mathrm{S}})$ monotonically decreases as $k\Delta_{\mathrm{S}}$ increases, the real and imaginary parts of $\mathrm{H}_0^{(2)}(k\Delta_{\mathrm{S}})$ oscillate with a period comparable to the wavelength. Therefore, complex interactions with the diagonal terms of $\mathbf{W}_{\mathrm{SS}}$ may give rise to the non-monotonic behavior of the linearity metric.
  \end{enumerate}

\begin{rem}

While the qualitative trends seen in Fig.~\ref{fig:RISfreespacePhysFad} are expected to be general, the quantitative values are specific to the RIS element properties. For example, one should not conclude based on Fig.~\ref{fig:RISfreespacePhysFad} that proximity-induced mutual coupling never yields $\langle\zeta\rangle$ below 10.8~dB for $\Delta_{\mathrm{S}}=0.5\lambda_0$; for more strongly scattering RIS elements, $\langle\zeta\rangle$ may well be much lower.
   
\end{rem}

\subsection{Experimental Analysis}
\label{freespaceRIS_experimental}

Our experimental RIS prototype is a $3 \times 5$ array of the RIS element presented in Ref.~\cite{kaina2014hybridized}. The spacing between the centers of neighboring RIS elements in both dimensions is 6~cm which is roughly half a wavelength for frequencies within its operating band centered on 2.45~GHz. 

\begin{rem}

It is possible to retrieve the polarizability of the experimentally used RIS element based on far-field measurements. The necessary procedure is detailed in Ref.~\cite{diebold2022patch}. Since our analysis and arguments do not require quantitatively precise knowledge of the RIS element's polarizability, such a characterization is left for future work.

\end{rem}

\begin{figure*}
    \centering
    \includegraphics[width=\columnwidth]{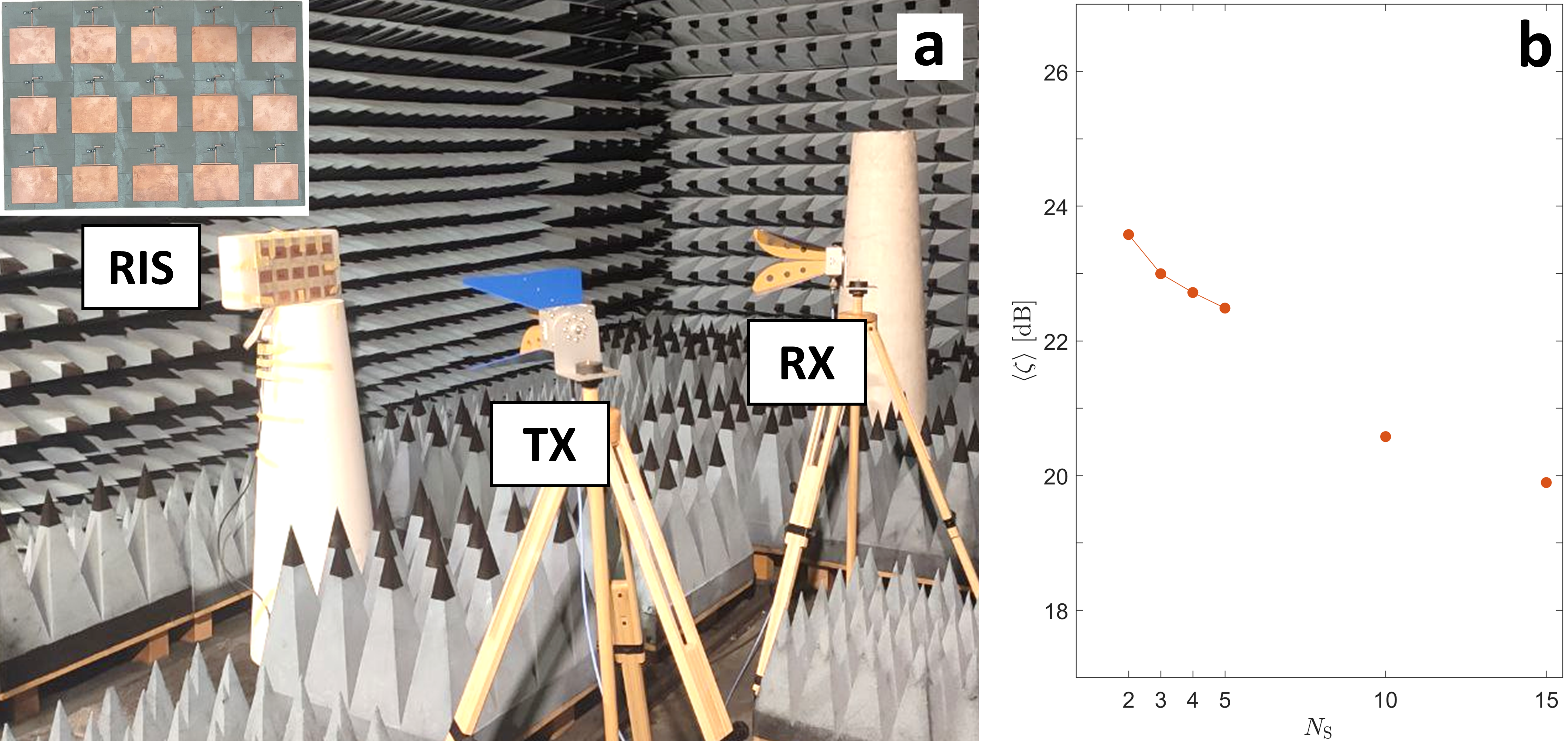}
    \caption{(a) Experimental setup in an anechoic (echo-free) chamber. The inset displays a frontal view of the RIS prototype. (b) Experimentally determined dependence of the linearity of a SISO channel's RIS parametrization in free space on the number of RIS elements. Every shown data point is averaged over 100 realizations with randomly chosen TX and RX positions.}
    \label{fig:RISfreespaceExp}
\end{figure*}

To mimic free space, we perform these experiments in an echo-free anechoic chamber. Its walls are fitted with absorbing material to prevent any reflections, such that operation in free space is emulated. Our experimental setup is shown in Fig.~\ref{fig:RISfreespaceExp}(a). We do not consider this anechoic chamber to be a realistic radio environment, but it is the environment that we need in order to isolate the effects of proximity-induced mutual coupling due to the RIS design. Incidentally, recent experimental studies of RIS-parametrized wireless channels such as Ref.~\cite{tang2020wireless} are limited to anechoic chambers as radio environment.

\begin{rem}
    We assume that our measurements are noise-free because we measure the wireless channels with a high-precision vector network analyzer (Rhode \& Schwarz ZVA 67). In addition, we choose to use horn antennas (Aaronia PowerLOG 70180) rather than omnidirectional dipole antennas in the experimental setup seen in Fig.~\ref{fig:RISfreespaceExp}(a) to improve the channel measurement's signal-to-noise ratio. In the present free-space scenario, under the assumption of noise-free measurements, the antenna's directivity does not impact the results because reflections in the anechoic environment can only originate from the RIS. Moreover, our linearity metric is insensitive to the RIS-independent constant component of the TX-RX wireless channel. 
    \label{remark_directivity_FreeSpace}
\end{rem}

We plot in Fig.~\ref{fig:RISfreespaceExp}(b) the experimentally determined dependence of the linearity metric on $N_{\mathrm{S}}$ for our RIS prototype operated in free space. Each data point is the average over 100 realizations of different relative positions of the TX, the RX and the RIS. To study values of $N_{\mathrm{S}}<15$ with our 15-element prototype, we pick $N_{\mathrm{S}}$ neighboring elements; we hold all remaining elements in a fixed configuration throughout such that they act just like a metal wall that does not impact our linearity metric because constant terms in the channel do not matter and there are no multi-path reflections in the anechoic environment. For $N_{\mathrm{S}}\in \{2,3,4,5\}$ we always pick horizontally neigboring elements whereas for $N_{\mathrm{S}}=10$ we pick two entire neighboring lines of five elements. Out of an abundance of caution, we hence present the data for $N_{\mathrm{S}}=10$ and $N_{\mathrm{S}}=15$ as individual dots in Fig.~\ref{fig:RISfreespaceExp}(b) since the mutual coupling effects might have been slightly different. In any case, the trend is very clear: the linearity metric monotonously deteriorates as $N_{\mathrm{S}}$ is increased, in line with our analytical and numerical findings from the previous subsections. 

\begin{rem}
    It is noteworthy that we experimentally detect the structural non-linearity even for free-space operation of our small-scale RIS prototype with half-wavelength spaced elements, i.e., under conditions that one would expect to be ideal for the applicability of the linear cascaded channel model.

\end{rem}

\section{RIS-Assisted MIMO Communications in Generic Radio Environments}
\label{sec:genericRIS}

Finally, we are now ready to consider RIS-assisted MIMO communications in generic radio environments that include scattering objects, unlike the trivial radio environment of free space considered in the previous Sec.~\ref{sec:freespaceRIS}. 
At microwave and millimeter-wave frequencies, which are important elements of 6G's all-spectra-integrated networks~\cite{you2021towards}, many important deployment scenarios such as factories~\cite{giordani2020toward} for machine-type communications will give rise to rich scattering~\cite{GeorgeMag}. Early experiments from 2016~\cite{del2016intensity} and 2018~\cite{del2018leveraging} conducted by one of the current authors (and coworkers) already noticed that describing RIS-parametrized wireless channels in rich-scattering environments with linear models was only approximately possible in the limiting case of strong absorption and hence short reverberation times.
This section contains again an analytical analysis (Sec.~\ref{rich_scat_analytical}), a numerical analysis  (Sec.~\ref{rich_scat_numerical}) and an experimental analysis  (Sec.~\ref{rich_scat_experimental}).

\subsection{Analytical Analysis}
\label{rich_scat_analytical}

For the most general problem of RIS-assisted MIMO communications in generic radio environments with arbitrarily complex scattering, $\mathbf{H} \propto [\mathbf{W}^{-1}]_{\mathrm{RT}}$ [see Fig.~\ref{fig:PhysFad_basics}(b)]. As in Fig.~\ref{fig:PhysFad_basics}(c), we begin by expressing $\mathbf{W}$ as $2 \times 2$ block matrix:

\begin{equation}
\mathbf{W} = 
    \begin{bmatrix} 
	\mathbf{W}_{\mathrm{TT}} & \mathbf{W}_{\mathrm{TR}} & \mathbf{W}_{\mathrm{TE}} & \mathbf{W}_{\mathrm{TS}} \\
	\mathbf{W}_{\mathrm{RT}} & \mathbf{W}_{\mathrm{RR}} & \mathbf{W}_{\mathrm{RE}} & \mathbf{W}_{\mathrm{RS}}\\
	\mathbf{W}_{\mathrm{ET}} & \mathbf{W}_{\mathrm{ER}} & \mathbf{W}_{\mathrm{EE}} & \mathbf{W}_{\mathrm{ES}}\\
 	\mathbf{W}_{\mathrm{ST}} & \mathbf{W}_{\mathrm{SR}} & \mathbf{W}_{\mathrm{SE}} & \mathbf{W}_{\mathrm{SS}}\\
	\end{bmatrix}
 =  \begin{bmatrix} 
	\mathbf{W}_3 & \mathbf{W}_{\mathrm{3S}} \\
	\mathbf{W}_{\mathrm{S3}}  & \mathbf{W}_{\mathrm{SS}} \\
\end{bmatrix},
 \label{eq:W}
\end{equation}

\noindent where $\mathbf{W}_{\mathrm{S3}} = [ \mathbf{W}_{\mathrm{ST}} \ \mathbf{W}_{\mathrm{SR}} \ \mathbf{W}_{\mathrm{SE}}]$ and  $\mathbf{W}_{\mathrm{3S}}= [ \mathbf{W}_{\mathrm{TS}} \ \mathbf{W}_{\mathrm{RS}}\ \mathbf{W}_{\mathrm{ES}}]^T$. We have not previously encountered $\mathbf{W}_{3}$ but its structure is mathematically analogous to that of $\mathbf{W}_{2}$ from Sec.~\ref{sec:mimo_freespace}; hence, upon replacing the RIS (``S'') by the radio environment (``E''), all results from Sec.~\ref{sec:mimo_freespace} can be directly applied to $\mathbf{W}_{3}$.

Following the same approach as in the previous sections, block-based matrix inversion yields
\begin{equation}
\begin{split}
    [\mathbf{W}^{-1}]_3  = \left( \mathbf{W}_3 - \mathbf{W}_{\mathrm{3S}} \mathbf{W}_{\mathrm{SS}}^{-1} \mathbf{W}_{\mathrm{S3}} \right)^{-1} 
    =    \mathbf{W}_{3}^{-1} \sum_{k=0}^\infty \left( \mathbf{W}_{\mathrm{3S}} \mathbf{W}_{\mathrm{SS}}^{-1} \mathbf{W}_{\mathrm{S3}} \mathbf{W}_{3}^{-1} \right)^k
    \\=\mathbf{W}_{3}^{-1} + \mathbf{W}_{3}^{-1}  \mathbf{W}_{\mathrm{3S}} \mathbf{W}_{\mathrm{SS}}^{-1} \mathbf{W}_{\mathrm{S3}} \mathbf{W}_{3}^{-1} + \mathbf{W}_{3}^{-1}  \left(\mathbf{W}_{\mathrm{3S}} \mathbf{W}_{\mathrm{SS}}^{-1} \mathbf{W}_{\mathrm{S3}} \mathbf{W}_{3}^{-1}\right)^2 + \dots
\end{split},
\label{eq_winv}
\end{equation}

\noindent The series converges if $\rho \left(\mathbf{W}_{\mathrm{3S}} \mathbf{W}_{\mathrm{SS}}^{-1} \mathbf{W}_{\mathrm{S3}}\mathbf{W}_{\mathrm{3}}^{-1}\right)<1$ and, once again, the norm of the common ratio of the Born-like series in Eq.~(\ref{eq_winv}) determines at which term the series can be truncated. This norm can be bounded as
\begin{equation}
    \lVert \mathbf{W}_{\mathrm{3S}} \mathbf{W}_{\mathrm{SS}}^{-1} \mathbf{W}_{\mathrm{S3}} \mathbf{W}_{3}^{-1} \rVert_2 \leq  \frac{|\alpha_\text{S}|}{1-\mathcal{C}_{\mathrm{S}} } \lVert \mathbf{W}_{\mathrm{3S}}\rVert_2^2 \lVert \mathbf{W}_{3}^{-1}\rVert_2,
\end{equation}
where we use $\lVert \mathbf{W}_{\mathrm{SS}}^{-1}\rVert_2 \leq \frac{|\alpha_\text{S}|}{1-\mathcal{C}_{\mathrm{S}} }$ from Sec.~\ref{sec:freespaceRIS}. Physically, the common ratio of the infinite series in Eq.~(\ref{eq_winv}) represents bounces from ``3'' (TX array, RX array and scattering environment) to the RIS and back to ``3''. The term $\frac{|\alpha_\text{S}|}{1-\mathcal{C}_{\mathrm{S}} }$ is proportional to the magnitude of the polarizability of the RIS elements, and does not vanish even if we assume zero proximity-induced mutual coupling between the RIS elements, i.e., $\mathcal{C}_\textrm{S}=0$. In other words, reverberation-induced long-range coupling is a mechanism giving rise to structural non-linearity that is independent of the previously identified mechanism due to proximity-induced mutual coupling. The term $\lVert \mathbf{W}_{\mathrm{3S}}\rVert_2$ depends on the distances between the RIS elements and the entities in ``3''. The term $\lVert \mathbf{W}_{3}^{-1}\rVert_2$ depends on the reverberation within ``3'' and depends on the number of antennas and scattering objects, their polarizabilities, and their proximity. 
``3'' is typically dominated by the scattering environment and here we \textit{cannot} argue that higher-order terms of the series are in general rapidly attenuated, mainly because the scattering strength of the environment may be substantial. 
For instance, under rich-scattering conditions, $\lVert\mathbf{W}_{\mathrm{EE}}^{-1}\rVert_2$ may be large which ultimately yields a large value of $\lVert[\mathbf{W}_{3}^{-1}]_{\mathrm{RT}}\rVert_2$, analogous to the dependence of $\lVert[\mathbf{W}_{2}^{-1}]_{\mathrm{RT}}\rVert_2$ on $\lVert\mathbf{W}_{\mathrm{SS}}^{-1}\rVert_2$ in Sec.~\ref{sec:mimo_freespace}, as we discuss below.

So far, we only worked out $[\mathbf{W}^{-1}]_3$ in Eq.~(\ref{eq_winv}) but we seek $\mathbf{H} \propto [\mathbf{W}^{-1}]_{\mathrm{RT}}$. By inserting the definition of $\mathbf{W}_{\mathrm{S3}}$ and $\mathbf{W}_{\mathrm{3S}}$ into Eq.~(\ref{eq_winv}), we can determine $[\mathbf{W}^{-1}]_{\mathrm{RT}}$ up to any desired order. However, for the present most general case, writing down all the terms becomes arduous even at second order; we trust that the reader has by now grasped the gist of how the series expansions work. Therefore, let us explicitly state that one can in general \textit{not} truncate the series expansion of $\mathbf{H}$ at linear order, and now let us do exactly that truncation at linear order anyway, purely for the purpose of discussing the physical meaning of the various resulting terms:

\begin{equation}
\mathbf{H} \propto [\mathbf{W}^{-1}]_{\mathrm{RT}} = [\mathbf{W}_{3}^{-1}]_{\mathrm{RT}} + [\mathbf{W}_3^{-1}]_{\mathrm{R-TRE}}  \mathbf{W}_{\mathrm{3S}} \mathbf{W}_{\mathrm{SS}}^{-1} \mathbf{W}_{\mathrm{S3}} [\mathbf{W}_3^{-1}]_{\mathrm{TRE-T}} + \dots
\label{born_full}
\end{equation}

\noindent where for compactness of notation we introduce $[\mathbf{W}_3^{-1}]_{\mathrm{R-TRE}} = [[\mathbf{W}_3^{-1}]_{\mathrm{RT}}  \ [\mathbf{W}_3^{-1}]_{\mathrm{RR}} \  [\mathbf{W}_3^{-1}]_{\mathrm{RE}}  ]$ and $[\mathbf{W}_3^{-1}]_{\mathrm{TRE-T}} = [[\mathbf{W}_3^{-1}]_{\mathrm{TT}}  \ [\mathbf{W}_3^{-1}]_{\mathrm{RT}} \  [\mathbf{W}_3^{-1}]_{\mathrm{ET}}  ]^T$. By evaluating the block vector products in Eq.~(\ref{born_full}), we can identify the following correspondences with the common cascaded model from Eq.~(\ref{eq:cascade}):

\begin{subequations}
\begin{align}
\mathbf{H_0} \longleftrightarrow   [\mathbf{W}_{3}^{-1}]_{\mathrm{RT}}\\
\mathbf{H_1} \longleftrightarrow   [\mathbf{W}_{3}^{-1}]_{\mathrm{RT}}\mathbf{W}_{\mathrm{TS}} + [\mathbf{W}_{3}^{-1}]_{\mathrm{RR}} \mathbf{W}_{\mathrm{RS}} + [\mathbf{W}_{3}^{-1}]_{\mathrm{RE}} \mathbf{W}_{\mathrm{ES}}\\
\mathbf{H_2} \longleftrightarrow   \mathbf{W}_{\mathrm{ST}}[\mathbf{W}_{3}^{-1}]_{\mathrm{TT}}  + \mathbf{W}_{\mathrm{SR}} [\mathbf{W}_{3}^{-1}]_{\mathrm{RT}} + \mathbf{W}_{\mathrm{SE}} [\mathbf{W}_{3}^{-1}]_{\mathrm{ET}}
\end{align}
\label{eq:interpretRISnENV}
\end{subequations}

\noindent The $\mathbf{H_0}$ term again captures all paths that do not involve any encounters with the RIS. Recall that by analogy (``S'' $\rightarrow$ ``E'' and $[\mathbf{W}_3^{-1}]_{\mathrm{RT}} \rightarrow [\mathbf{W}_2^{-1}]_{\mathrm{RT}}$) we have already worked out an expression for $[\mathbf{W}_3^{-1}]_{\mathrm{RT}}$ in Sec.~\ref{sec:freespaceRIS}. Recall also that this expression is itself an infinite series representing scattering between the TX, the RX and the scattering environment. For strongly scattering radio environments with strong $\lVert\mathbf{W}_{\mathrm{EE}}^{-1}\rVert_2$, many terms of this series may be significant.

\begin{rem}

The constant term $\mathbf{H_0}$ can in general not be identified simply as the LOS path between TX and RX. Instead, besides the LOS path, $\mathbf{H_0}$ includes all multi-bounce paths that never encounter the RIS. 

\end{rem}

\noindent The $\mathbf{H_1}$ term contains the three families of possible paths from the RIS to the RX array without any additional encounters with the RIS along the way, namely those that start with a trajectory from the RIS to the TX array, the RX array or the environment, and eventually arrive at the RX array without revisiting the RIS.
Similarly, the $\mathbf{H_2}$ term contains the three possible families of paths from the TX array to the RIS without any previous encounters with the RIS along the way, namely those that start from the transmitter, and, without any previous encounters with the RIS, finish with a trajectory from the TX array, the RX array or the environment to the RIS.

It is clear how this expansion will continue for higher-order terms, including at the $K$th order all those paths that encountered the RIS $K$ times. Truncating the Born-like series from Eq.~(\ref{born_full}) at linear order, and hence neglecting any terms depending non-linearly on $\mathbf{W}_{\mathrm{SS}}^{-1}$, is justified if:
\begin{itemize}
  \item the RIS is very far away from the TX array, the RX array \textit{and} any scattering object that is part of the radio environment (leading to very small $\lVert  \mathbf{W}_{\mathrm{3S}}\rVert_2 = \lVert  \mathbf{W}_{\mathrm{S3}}\rVert_2$).
\end{itemize}

\noindent and/or 

\begin{itemize}
\item the scattering cross-section of the TX array, the RX array \textit{and} all scattering objects constituting the radio environment is very small (leading to a very small $\lVert \mathbf{W}_3^{-1} \rVert_2$).
\end{itemize}

\noindent If the radio environment is just free space, these conditions collapse to Assumption~2. However, for a complex scattering environment, it will in general \textit{not} be justified to truncate the Born-like series from Eq.~(\ref{born_full}) at linear order because the environment's scattering cross-section is large rather than negligible. Hence, terms that are non-linear in $\mathbf{W}_{\mathrm{SS}}^{-1}$ will in general play a significant role. Then, even if Assumption~3 holds (i.e., if $\mathbf{\mathcal{M}}_{\mathrm{SS}}^{-1} = \mathbf{0}$ such that $\mathbf{W}_{\mathrm{SS}}^{-1} = \mathbf{\Phi})$, the wireless channel will depend in a non-linear manner on the RIS configuration due to reverberation-induced long-range coupling. 

Reverberation-induced long-range coupling is a source of structural non-linearity that is completely independent from the proximity-induced mutual coupling that we discussed in Sec.~\ref{sec:freespaceRIS}, and its underlying mathematical origin is different. Based on our analysis, we can make the following physically intuitive observations about what parameters will determine the importance of reverberation-induced long-range correlations:

\begin{enumerate}
    \item \textit{Reverberation time.} The longer the wave reverberates in the radio environment, the higher is the probability that a given path encounters the RIS multiple times. To be more precise, the wave gradually vanishes due to attenuation and leakage. In a rich-scattering environment, the field intensity decays exponentially on average with a decay constant $1/\tau$~\cite{west2017best}. Naturally, the decay of a typical rich-scattering channel impulse response is of the same exponential nature. To quantify the reverberation time, one can hence use $\tau$ (extracted from the average measured channel impulse response decay rate), considering that paths taking longer than $\tau$ to reach the receiver are so heavily attenuated that they do not significantly contribute to the received signal. Translating the reverberation time into a number of bounces $\mathcal{N}$ is not straightforward. For the case of a roughly cubic enclosure of volume $V$ as radio environment (e.g., inside a room or vessel), the mean free path between scattering events is on the order of $\sqrt[3]{V}$ and hence a rough estimate is $\mathcal{N} = \frac{\tau c}{\sqrt[3]{V}}$, where $c$ is the speed of light.
    
    \begin{rem}
        $\mathcal{N}$ is a loose upper bound rather than a useful estimate of the order $K$ at which we can truncate the Born-like series from Eq.~(\ref{eq_winv}) with acceptable loss of precision because we have not yet considered how many of the $\mathcal{N}$ bounces involve the RIS. This depends on the dominance of the RIS in the radio environment (see below).
    \end{rem}

    \item \textit{Dominance of RIS in radio environment.} The likeliness that a given bounce involves the RIS depends on the dominance of the RIS: the larger the RIS and the stronger the scattering strength of the RIS elements is, the more dominant the RIS will be and the stronger will be the structural non-linearity due to reverberation-induced long-range coupling. Other factors like the relative positions of the wireless entities also play a role but are very difficult to quantify. If the RIS elements were randomly distributed across the radio environment's walls, one could estimate that an acceptable truncation of the Born-like series would be at order $K\approx \frac{N_{\mathrm{S}}\sigma_{\mathrm{S}}}{A_{\mathrm{E}}}\mathcal{N}$, where $\sigma_{\mathrm{S}}$ is the scattering cross-section of an individual RIS element and $A_{\mathrm{E}}$ denotes the radio environment's surface area.

\end{enumerate}

\begin{rem}
    Since the radio environment (excluding the RIS) is not under the wireless system engineer's control, there is no possibility to reduce the importance of reverberation-induced long-range coupling in a given radio environment.
\end{rem}

\subsection{Numerical Analysis}
\label{rich_scat_numerical}

To confirm these analytical insights into the extent to which reverberation-induced long-range coupling between the RIS elements limits the applicability of the linear cascaded channel model in generic (possibly rich-scattering) radio environments, we now conduct numerical tests with PhysFad~\cite{PhysFad}. We consider again a RIS-assisted SISO system, but this time in a complex scattering environment made up of a dipole fence constituting an electrically-large irregularly shaped enclosure as well as some dipoles inside the enclosure. By varying the resonance frequency $f_{\mathrm{res}}^{\mathrm{E}}$ of the dipoles representing the scattering environment, we control their scattering strength (see Ref.~\cite{PhysFad} for details). As $f_{\mathrm{res}}^{\mathrm{E}}$ gets larger, the scattering environment becomes gradually more transparent and in the limit of $f_{\mathrm{res}}^{\mathrm{E}}\rightarrow \infty$ we recover a free-space radio environment. 

\begin{figure*}
    \centering
    \includegraphics[width=\columnwidth]{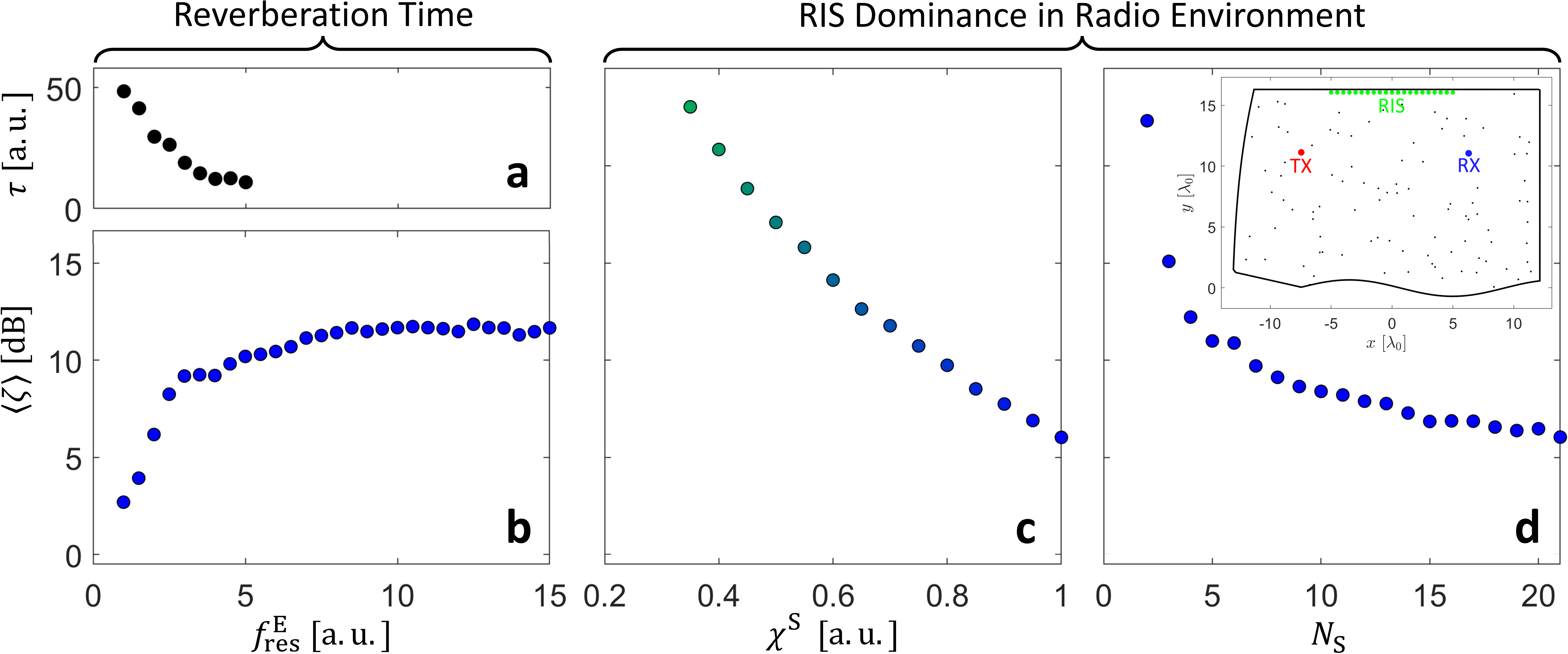}
    \caption{Numerical evaluation based on PhysFad of how the linearity of a SISO channel's RIS parametrization in the radio environment with complex scattering (sketched in the inset) depends on $f_{\mathrm{res}}^{\mathrm{E}}$ (b) [the reverberation time depends directly on $f_{\mathrm{res}}^{\mathrm{E}}$ (a)], the RIS element's scattering strength (c), and the number of RIS elements (d). Every shown data point is averaged over 100 realizations with randomly chosen TX position (red) and RX position (blue) within the enclosure. The default parameters (all but the one plotted on the horizontal axis of a given subplot) are $f_{\mathrm{res}}^{\mathrm{E}}=2 \ \mathrm{a.u.}$, $\chi^{\mathrm{S}}=1\ \mathrm{a.u.}$, $N_{\mathrm{S}}=21$ and $\Delta_{\mathrm{S}}=0.5\lambda_0$.}
    \label{fig:RISrichscattPhysFad}
\end{figure*}

We observe in Fig.~\ref{fig:RISrichscattPhysFad}(a) that indeed the reverberation time $\tau$ rapidly decreases as $f_{\mathrm{res}}^{\mathrm{E}}$ is increased because the dipoles constituting the scattering environment become increasingly transparent. For $f_{\mathrm{res}}^{\mathrm{E}}>5$ we can no longer extract an exponential decay constant based on the channel impulse response because the amount of scattering has become (almost) negligible. The error due to assuming that the wireless channel linearly depends on the RIS configuration is very strong under rich-scattering conditions ($\langle\zeta\rangle = 2.6$~dB in our setting which is not extremely reverberant) but rapidly decreases as the reverberation time decreases, and our linearity metric converges to its corresponding free-space value from Fig.~\ref{fig:RISfreespacePhysFad} for $f_{\mathrm{res}}^{\mathrm{E}}>5$. 

We also observe that the linearity metric is higher if the RIS is less dominant, meaning that it is less likely that a given path involves multiple bounces off the RIS. Specifically, Fig.~\ref{fig:RISrichscattPhysFad}(c) evidences that the smaller each RIS element's scattering strength is (controlled via $\chi^{\mathrm{S}}$), the higher is the linearity metric. Recall, however, that we seek RIS elements with strong scattering cross-section that significantly impact the wireless channel. Moreover, we see in Fig.~\ref{fig:RISrichscattPhysFad}(d) that the more RIS elements there are, the lower is the linearity metric.

\subsection{Experimental Analysis}
\label{rich_scat_experimental}

\begin{figure*}[h]
    \centering
    \includegraphics[width=\columnwidth]{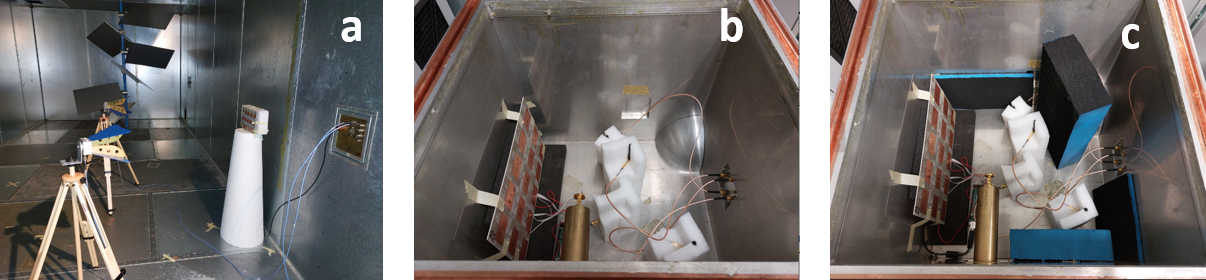}
    \caption{Photographic images of some experimental setups used to measure the linearity metric under rich-scattering conditions. See Table~I for details.}
    \label{fig:RISrichscattExp}
\end{figure*}

We now experimentally explore using our 15-element RIS prototype how the linearity metric varies in rich-scattering environments depending on the utilized antenna type, percentage of surface covered by the RIS, and the reverberation time. The experimental setups are depicted in Fig.~\ref{fig:RISrichscattExp} and key results are summarized in Table~I.

\vspace{-0.95cm}

\begin{table}[h]

\caption{Summary of key characteristics and determined  linearity metric in various experimentally studied complex scattering scenarios. The linearity metrics are measured for the case of using all $N_{\mathrm{S}}=15$ elements of our RIS prototype. (AC: anechoic chamber; RC: reverberation chamber).}

\label{tbl:Symbols}
\begin{center}
\begin{tabular}{ |c|c|c|c|c|c|c|c| } 
\hline
Figure & Antenna & Radio Environment & Volume~[$\mathrm{m}^3$] & Surface~[$\mathrm{m}^2$]   & $Q$\footnotemark & $\tau\ [\mathrm{ns}]$ & $\langle\zeta\rangle$~[dB] \\

\hline

 Fig.~\ref{fig:RISfreespaceExp}(a) & horn & Free Space (AC) & N/A & N/A &   N/A & N/A & 19.9 \\

Fig.~\ref{fig:RISrichscattExp}(a) & horn & Large RC & 93.4 & 136.3 & 42651 & $2.8\times10^3$ & 12.6 \\

N/A  & dipole & Large RC &   93.4 & 136.3 & 42651 & $2.8\times10^3$ & 12.1 \\

Fig.~\ref{fig:RISrichscattExp}(b) & dipole &  Tiny RC &   0.07 & 0.98 & 475 & 30.8 & 7.9 \\

N/A  & dipole & Tiny RC w/ absorb. I &   0.07 & 0.98 & 256 & 16.6 & 9.4 \\

Fig.~\ref{fig:RISrichscattExp}(c)  & dipole & Tiny RC w/ absorb. II &    0.07 & 0.98 & 70 & 4.5 & 13.4 \\

N/A  & dipole & Tiny RC w/ absorb. III &  0.07 & 0.98 & 58 & 3.7 & 15.0 \\

\hline
\end{tabular}
\end{center}
\end{table}\label{TableRichScattExp}

\footnotetext[4]{The composite quality factor $Q$ of the radio environment is a common dimensionless metric in electromagnetic compatibility to quantify how quickly the energy of the electromagnetic wave field decays~\cite{west2017best}.}

By far the highest linearity metric is measured in free space. All cases involving complex scattering environments yield lower linearity metrics. First, we use a large reverberation chamber as radio environment with an extremely long reverberation time on the order of milliseconds, but the RIS only covers a tiny portion of the overall surface. The linearity metric is around 12~dB and hence notably lower than in free space. Despite the extremely long reverberation time, the linearity metric is not extremely low because the RIS does not dominate the radio environment due to its comparatively tiny size. We also observe that the linearity metric is slightly higher if we use horn antennas rather than dipole antennas.

\begin{rem}
Unlike in the free-space case from Sec.~\ref{sec:freespaceRIS} (cf. Remark~\ref{remark_directivity_FreeSpace}), the antenna directivity does directly impact the linearity metric for operation in a generic scattering environment: directive antennas pointed at the RIS preferentially capture the TX-RIS-RX path and reduce the probability that multi-bounce paths (which may encounter the RIS multiple times) are captured. The two entries  in Table~I corresponding to the large reverberation chamber that only differ in the type of antenna confirm this expectation. 
\label{remark_directivity_RichScatt}
\end{rem}

In order to examine a setting in which the dominance of the RIS is significant, we also performed measurements in a tiny reverberation chamber. However, its reverberation time is two orders of magnitude smaller. Nonetheless, we measure a low linearity metric of roughly 8~dB, evidencing that despite the lower reverberation time, the probability that rays encounter the RIS more than once is much higher in the tiny reverberation chamber. Finally, we now purposefully reduce its reverberation time by adding pieces of absorbing material. Table~I evidences that as we add absorbing material, the reverberation time drops and the linearity metric increases.

Overall, we hence report in the present section clear experimental evidence for the existence of a non-linear dependence of the wireless channel on the RIS configuration that originates from reverberation. Indeed, since we always use the same RIS prototype, the proximity-induced mutual coupling is the same in all cases listed in Table~I and hence cannot explain the measured differences in the linearity metric. Moreover, compared to Fig.~\ref{fig:RISfreespaceExp}, we see that reverberation-induced long-range coupling may contribute substantially more structural non-linearity than proximity-induced mutual coupling.

\section{Conclusions}\label{sec:Conclusion}

To summarize, through analytical, numerical and experimental analysis we have identified two distinct mechanisms that can give rise to a non-linear relation between the RIS configuration and the RIS-dependent wireless channel: \textit{i)} proximity-induced mutual coupling, and \textit{ii)} reverberation-induced long-range coupling. 
Of these two mechanisms, only the first one was previously documented (although no details of its non-linearity had been worked out).
While the second mechanism is an uncontrollable property of the radio environment, non-linear effects due to the first mechanism can be limited (or even mitigated) through careful design of the RIS hardware. 
Looking forward, this insight identifies an important avenue for future research in RIS hardware design: deliberate efforts are needed to minimize proximity-induced mutual coupling. This can be achieved in particular by including decoupling mechanisms which are already established for conventional patch-antenna arrays~\cite{wu2017array,vishvaksenan2017mutual,li2018isolation,lin2020weak,zhang2021novel,zhang2021simple}. 
Moreover, our new insights into the physics underlying the mathematical machinery involved in PhysFad~\cite{PhysFad} will help to identify methods to substantially speed up the required matrix inversion.

Common cascaded models of RIS-parametrized channels neglect both structural non-linearity mechanisms by tacitly assuming that the wireless channel depends linearly on the RIS configuration. We have shown that this assumption is equivalent to a truncation of a Born series (first mechanism) and a Born-like series (second mechanism) after the first and second term, respectively. Especially in rich-scattering environments, the latter truncation will cause significant errors due to reverberation-induced long-range coupling -- even if proximity-induced mutual coupling is negligible thanks to carefully designed RIS hardware. Enriching the common cascaded model with higher-order (multi-bounce) terms is in principle feasible but appears cumbersome and difficult to manipulate in signal processing. In contrast, PhysFad~\cite{PhysFad} compactly captures all terms of the infinite series through a matrix inversion and thereby constitutes a physics-compliant end-to-end channel model for arbitrarily complex generic RIS-parametrized radio environments.

We have also worked out and evidenced which factors determine the importance of the two mechanisms that give rise to structural non-linearity. Proximity-induced mutual coupling depends on the scattering strength, number and spatial arrangement of the RIS elements; its dependence on the spatial arrangement promises to be a powerful tuning knob to mitigate proximity-induced mutual coupling, as stated above. Reverberation-induced long-range coupling depends on the probability that a path from the TX to the RX encounters the RIS multiple times, which in turn depends on the reverberation time, the percentage of the environment's surface that is covered by the RIS, as well as the relative location of the RIS with respect to the other wireless entities. Operating with large RISs in strongly scattering environments (factories, vessels, etc.) likely corresponds to a regime in which a linear cascaded channel model performs quite poorly. Measurement campaigns to rigorously analyze realistic RIS-parametrized rich-scattering environments constitute an important direction for future research.

The to-date largely neglected structural non-linearity of RIS-parametrized radio environments has two important consequences whenever the linear cascaded approximation becomes inaccurate:
\begin{itemize}
    \item 
First, the reliability of existing performance predictions based on unjustified linearity assumptions, especially for rich-scattering radio-environments, must be questioned~\cite{AntoninEuCAP}. Linear models may drastically undererstimate the potential of RIS-based wave control. For instance, wave-based signal processing operations can be implemented much more precisely and flexibly with RISs in rich-scattering conditions than in their free-space counterpart~\cite{sol2022meta}. The reason is that strongly reverberating wave fields are much more sensitive to perturbations such as the RIS configuration. The same argument also explains why the achievable localization precision is drastically enhanced under rich-scattering conditions~\cite{del2021deeply}. 
\item Second, the channel-estimation procedures that are currently being developed based on the cascade assumption cannot be applied in realistic settings where the linearity assumption cannot be justified. Instead, in rich-scattering radio environments, the acquisition of full context-awareness will be required in order to predict the end-to-end channel for a given RIS configuration, which will give rise to a need for integrated sensing and communications (ISAC) for the operation of self-adaptive RISs under rich-scattering conditions, as recently pointed out by one of the current authors and a coworker in Ref.~\cite{ChloeMag}.
\end{itemize}

\section*{Acknowledgment}

P.d.H. thanks I.~Ahmed for his precious help in fabricating the RIS prototype.

\appendices

\section{Derivation of Eq.~(\ref{definition_C}) and Eq.~(\ref{bound_antenna_array})}\label{AppendixA}

Using the triangle inequality and the submultiplicativity of the matrix 2-norm, we find that  $\left\Vert\mathbf{W}_{\mathrm{TT}}^{-1}\right\Vert_2 \leq \left\Vert\mathbf{\Omega}_{\mathrm{TT}}\right\Vert_2 \sum_{k=0}^\infty \left\Vert(\mathbf{\Omega}_{\mathrm{TT}} \mathbf{\mathcal{M}}_{\mathrm{TT}})^k \right\Vert_2$.
Assuming identical antennas with polarizability $\alpha_\mathrm{T}$, it follows that $\left\Vert\mathbf{\Omega}_{\mathrm{TT}}\right\Vert_2 = |\alpha_\mathrm{T}|$ and we can identify the matrix $\mathbf{\Omega}_{\mathrm{TT}} \mathbf{\mathcal{M}}_{\mathrm{TT}}$ as being complex symmetric and hollow (all diagonal entries are zeros).

\begin{lem}
\label{lem:norm_comp_sym}
If $\mathbf{A}$ is a complex symmetric hollow matrix, then $\left\Vert \mathbf{A} \right\Vert_2 \leq \max_i\sum\nolimits_{j\neq i}|a_{ij}|$.
\end{lem}

\noindent By applying Lemma~\ref{lem:norm_comp_sym} (which we prove in Appendix~\ref{AppendixC}) to the matrix $\mathbf{\Omega}_{\mathrm{TT}} \mathbf{\mathcal{M}}_{\mathrm{TT}}$, we obtain $\left\Vert\mathbf{\Omega}_{\mathrm{TT}} \mathbf{\mathcal{M}}_{\mathrm{TT}} \right\Vert_2 \leq |\alpha_\text{T}|\ \underset{i \in [1,N_\text{T}]}{\text{max}}\,  \sum_{\substack{j\in [1,N_\text{T}] \\ j \neq i}} |G_{ij}|$ and hence $\left\Vert\mathbf{W}_{\mathrm{TT}}^{-1}\right\Vert_2 \leq |\alpha_\mathrm{T}| \sum_{k=0}^\infty \mathcal{C_\mathrm{T}}^k$. If $\mathcal{C_\mathrm{T}}<1$, this geometric series converges and we get $\left\Vert\mathbf{W}_{\mathrm{TT}}^{-1}\right\Vert_2 \leq  \frac{|\alpha_\mathrm{T}|}{1-\mathcal{C_\mathrm{T}}}$. We have hence derived Eq.~(\ref{definition_C}).

The error due to truncating the series after $K$ terms is $\mathbf{E}_K \triangleq \left(\sum_{k=K}^\infty
\left( -\mathbf{\Omega}_{\mathrm{TT}} \mathbf{\mathcal{M}}_{\mathrm{TT}} \right)^{k}\right) \mathbf{\Omega}_{\mathrm{TT}}$. For instance, the linear cascaded model assumes a truncation after the first term that yields the error $\mathbf{E}_1  =-\mathbf{\Omega}_{\mathrm{TT}} \mathbf{\mathcal{M}}_{\mathrm{TT}}\mathbf{W}_{\mathrm{TT}}^{-1}$ with $\left\Vert \mathbf{E}_1 \right\Vert_2 \leq \mathcal{C}_\mathrm{T} \left\Vert\mathbf{W}_{\mathrm{TT}}^{-1}\right\Vert_2 \leq \frac{|\alpha_\mathrm{T}|\mathcal{C}_\mathrm{T}}{1-\mathcal{C}_\mathrm{T}}$. More generally, we have $\mathbf{E}_K  =(-\mathbf{\Omega}_{\mathrm{TT}} \mathbf{\mathcal{M}}_{\mathrm{TT}})^K\mathbf{W_{\mathrm{TT}}}^{-1}$ and $\left\Vert \mathbf{E}_K \right\Vert_2  \leq \frac{|\alpha_\mathrm{T}|\mathcal{C}_\mathrm{T}^K}{1-\mathcal{C}_\mathrm{T}}$. If we seek to bound the relative rather than absolute truncation error, we must first bound $\mathbf{W}_{\mathrm{TT}}$ by applying the triangle inequality to Eq.~(\ref{eq8}): $
\left\Vert \mathbf{W}_{\mathrm{TT}} \right\Vert_2 \leq  \left\Vert \mathbf{\Omega}_{\mathrm{TT}}^{-1} \right\Vert_2 +\left\Vert \mathbf{\mathcal{M}}_{\mathrm{TT}} \right\Vert_2 \leq \frac{1}{|\alpha_\mathrm{T}|} + \frac{\mathcal{C}_\mathrm{T}}{|\alpha_\mathrm{T}|},
$. Then, we find that the normalized mean square error due to truncating the Born series after $K$ terms is $\left\Vert \mathbf{E}_K \right\Vert_2 \left\Vert \mathbf{W}_{\mathrm{TT}} \right\Vert_2 \leq \mathcal{C}_\mathrm{T}^K\frac{1+\mathcal{C}_\mathrm{T}}{1-\mathcal{C}_\mathrm{T}}$, as stated in Eq.~(\ref{bound_antenna_array}).

\section{Derivation of Eq.~(\ref{eq:WTRRRRTTT_bound})}\label{AppendixB}

Our goal is to bound the norm of the common ratio of the Born-like series in Eq.~(\ref{eq_w1inv}), $\left\Vert\mathbf{W}_{\mathrm{TR}} \mathbf{W}_{\mathrm{RR}}^{-1} \mathbf{W}_{\mathrm{RT}} \mathbf{W}_{\mathrm{TT}}^{-1}\right\Vert_2$. First, note that $\left\Vert\mathbf{W}_{\mathrm{TR}}\right\Vert_2 = \left\Vert\mathbf{W}_{\mathrm{RT}}\right\Vert_2 \leq \sqrt{N_{\text{T}}N_{\text{R}}} \frac{k^2}{4\epsilon\delta} \left| \text{H}_0^{(2)}\left(kD_{\text{RT}}\right) \right|$, where $D_{\text{RT}}$ denotes the smallest distance between any transmitter-receiver antenna pair. This bound is attained if the distances between all TX-RX antenna pairs are equal, making $\mathbf{W}_{\mathrm{TR}}$ a rank-1 matrix which corresponds to the plane-wave assumption commonly used in MIMO signal processing. Second, recall that we derived a bound on $\left\Vert\mathbf{W}_{\mathrm{TT}}^{-1}\right\Vert_2$ in Sec.~\ref{sec0}, which applies by analogy also to $\left\Vert\mathbf{W}_{\mathrm{RR}}^{-1}\right\Vert_2$. Now, by invoking the triangle inequality and the submultiplicativity of the matrix $2$-norm, we directly obtain Eq.~(\ref{eq:WTRRRRTTT_bound}).
This bound is $O(\frac{1}{D_{\text{RT}}})$.

\section{Proof of lemma~\ref{lem:norm_comp_sym}}\label{AppendixC}

\begin{proof}
Let $\mathbf{A}$ be a complex symmetric matrix, then a direct corollary of Ref.~\cite[Theorem~II]{takagi1924on} (explicitly stated in Ref.~\cite{garcia2007complex}) is that its 2-norm is
$
\left\Vert \mathbf{A} \right\Vert_2 = \sup \left\{\lambda>0 \,|\, \exists \mathbf{x}\neq 0,\, \mathbf{Ax} = \lambda \mathbf{x}^*\right\}.
$ 
We now prove an equivalent of the Gershgorin theorem~\cite{gershgorin1931uber} for this antilinear eigenvalue problem. With $x_i = |x_i|\mathrm{e}^{\mathrm{j}\phi_i}$, we get
\begin{equation}
\begin{split}
\mathbf{Ax} = \lambda \mathbf{x}^*  & \Leftrightarrow \forall i,\, \lambda x_i^* = \sum\nolimits_{j}a_{ij}x_j \\
& \Leftrightarrow \forall i,\, (\lambda x_i^*-a_{ii}x_i) = \sum\nolimits_{j\neq i}a_{ij}x_j \\
& \Rightarrow \forall i,\, \left|\lambda \mathrm{e}^{-\mathrm{j}\phi_i}-a_{ii}\mathrm{e}^{\mathrm{j}\phi_i}\right||x_i| = \left| \sum\nolimits_{j\neq i}a_{ij}x_j\right| \\ 
& \Rightarrow \forall i,\, \big|\lambda-|a_{ii}|\big| \leq \sum\nolimits_{j\neq i}|a_{ij}|\frac{|x_j|}{|x_i|} \\ 
& \Rightarrow \exists i,\, \big|\lambda-|a_{ii}|\big| \leq \sum\nolimits_{j\neq i}|a_{ij}| \\ 
\end{split}
\end{equation}
If $\mathbf{A}$ is a hollow matrix, $|a_{ii}|=0$, which yields $\lambda\leq \max_i\sum\nolimits_{j\neq i}|a_{ij}|$, such that we directly obtain
$
\left\Vert \mathbf{A} \right\Vert_2 \leq \max_i\sum\nolimits_{j\neq i}|a_{ij}|.
$
\end{proof}

\bibliographystyle{IEEEtran}


\end{document}